\documentclass[aps,prb,superscriptaddress,reprint,showpacs,floatfix]{revtex4-2}
\usepackage{graphicx}% Include figure files
\usepackage{dcolumn}% Align table columns on decimal point
\usepackage{bm}% bold math
\usepackage[dvipsnames]{xcolor}
\usepackage{relsize}
\usepackage{amsmath}
\usepackage{amsfonts}
\usepackage{amssymb}
\usepackage{mathtools}
\usepackage{braket}
\usepackage{gensymb}
\usepackage{url}
\usepackage[colorlinks = true,
            linkcolor = blue,
            urlcolor  = blue,
            citecolor = blue,
            anchorcolor = blue]{hyperref}
\usepackage{footmisc}
\usepackage[utf8]{inputenc}
\usepackage[T1]{fontenc}

\begin{document}

\preprint{APS/123-QED}

\title{Spin-Meissner effect in systems of coupled polariton condensates}

\author{I. Yu. Chestnov}
\affiliation{School of Physics and Engineering, ITMO University, St. Petersburg 197101, Russia}

\author{A. Kudlis}
\affiliation{Science Institute, University of Iceland, Dunhagi 3, IS-107 Reykjavik, Iceland}

\author{A. V. Nalitov}
\affiliation{Abrikosov Center for Theoretical Physics, MIPT, Dolgoprudnyi, Moscow Region 141701, Russia}

\author{I. A. Shelykh}
\affiliation{Science Institute, University of Iceland, Dunhagi 3, IS-107 Reykjavik, Iceland}

\date{\today}

\begin{abstract}
We theoretically investigate the interplay between Zeeman splitting and TE-TM-induced spin-flip tunneling in coupled exciton-polariton condensates systems and its impact on the spin–Meissner effect. We demonstrate that although a single condensate exhibits the effect of full paramagnetic screening via spin-anisotropic interactions, the inter-site spin-flip tunneling can dramatically alter this behavior. The geometry of the system is shown to play a crucial role.  In particular, in a dyad, the chemical potential reveals quadratic scaling with the magnetic field. In a triangle, the competition between Zeeman and TE-TM splittings produces a rich phase diagram that features asymmetric polarization states corresponding to both positive and negative magnetic susceptibility. In a square configuration, the symmetry of the network can restore the spin-Meissner effect, so that the condensate emission frequency becomes magnetic field independent in an extended parameter range. These findings not only shed light on the fundamental physics of polariton lattices but also suggest promising avenues for engineering robust spin–controlled photonic devices and polaritonic simulators.
\end{abstract}

\maketitle

\section{Introduction}

The studies of lattice systems cover an important part of modern physics, ranging from the conventional theory of solids to quantum magnetism \cite{Balents2010} and high-temperature superconductivity \cite{PatrickLee2006}. Developing a wide variety of emergent phases and associated phase transitions, lattice models also play an important role in the engineering of novel materials with predicted properties \cite{Das2013}.
An important tool for the exploration and simulation of lattice systems was gained after the experimental realization of cold atomic condensates in optical lattices \cite{Bloch2008}. 
The possibility of precise control of the parameters of this system made possible the study of such phenomena as the superfluid to Mott insulator phase transition \cite{Greiner2002}, realization of the Tonks-Girardeau bosonic gas \cite{Kinoshita2004,Paredes2004}, Bloch oscillations \cite{Morsch2001} and quantum simulation of magnetic systems \cite{Simon2011}. However, while being a perfect playground for theorists, cold atomic systems typically require ultra-low temperature ($< 1~\mu$K), which severely limits their possible applications.

Another field with prominent advances in the exploration of lattice systems is mesoscopic nonlinear optics.
Various optical lattices formed by coupled optical resonators were successfully realized, giving rise to observation of such phenomena as optical Bloch oscillations \cite{Trompeter2006}, photonic topological insulators \cite{Khanikaev2013,Rechtsman2013,Liang2013}, optical gap solitons \cite{Fleischer2003}, vortices \cite{Li2023} {\it etc}. 
In addition, a large number of effects was theoretically predicted for these systems, including optical fractional Hall effect \cite{Umucalilar2012}, topological delay lines \cite{Hafezi2011}, superfluid-Mott insulator optical phase transition \cite{Greentree2006}, and others. 
The major evident advantage in comparison to cold atomic systems is the possibility of operating at substantially higher (up to room) temperature. 
At the same time, the operability of optical lattice systems is restricted by a typically small value of nonlinearities.

One of the efficient ways to achieve strong optical nonlinearity is to use polaritonic systems enabled by the strong coupling regime between photons and excitons, achieved in optical microcavities \cite{Kavokin2017_OxfPr,Carusotto2013}.
Exciton polaritons, hybrid half-light half-matter particles, inherit extremely small effective mass (about $10^{-5}$ of the mass of free electrons) and large coherence length (on the micrometer scale) from their photonic component \cite{Ballarini2017}. At the same time, the excitonic component provides efficient polariton-polariton interactions which lead to extremely strong nonlinear optical response.
Polariton-based structures have been shown to be extremely promising for the observation of quantum coherent phenomena including BEC and superfluidity at surprisingly high temperatures under both optical \cite{Kasprzak2006,Balili2007} and electrical \cite{Schneider2013} excitation.

Recent technological advances allowed routine engineering of polaritonic lattices \cite{PhysRevLett.116.066402,Milicevic2017,Gao2018,Whittaker2018,Whittaker2021}. Combining the properties of cold atomic condensates, where macroscopic coherence plays an important role, with accessibility and controllability of optical systems, polaritonic setups are extremely promising candidates for the realization of novel types of optical metamaterials \cite{Kozin2018}. The possibility of using polariton lattices in quantum computing \cite{Ghosh2020_QI,Kavokin2022} and as quantum and neuromorphic simulators was also actively discussed \cite{Kavokin2022,Kalinin2018,Suchomel2018,Opala2019,Boulier2020_AQT,Marcucci2020}.

An important property of cavity polaritons is their spin degree of freedom \cite{ShelykhReview} inherited from the spins of quantum well excitons and cavity photons.
Similarly to photons, polaritons have two possible spin projections on the structure growth axis, corresponding to the two opposite circular polarizations, which can be mixed by effective magnetic fields of various origins.
External magnetic fields, applied along the structure growth axis and acting on the excitonic component via the Zeeman effect, lift the degeneracy of polariton states with opposite circular polarizations, while TE-TM energy splitting of the photonic modes in a planar cavity couples these states to each other via a $k$-dependent term, thus playing the role of an effective spin-orbit interaction \cite{ShelykhReview,Sala2015}.
In samples, which are patterned to introduce a lattice potential, the combination of the Zeeman and TE-TM splittings can be used to open topological gaps in honeycomb ~\cite{Nalitov2015,klembt2018exciton} and kagome lattices \cite{Gulevich-kagome}.

%tell something about requirement of equilibrium

Importantly, polariton-polariton interactions are also spin-dependent \cite{Vladimirova2010}.
Indeed, they stem from the interactions of excitons which is dominated by the exchange mechanism \cite{Ciuti1998}.
As a result, polaritons with aligned spins interact orders of magnitude stronger than polaritons with opposite circular polarizations \cite{Glazov2009}.
This property leads to the effect of self-induced Larmor precession of polariton pseudospins in an effective magnetic field \cite{Shelykh2004,Laussy2006}.
In the presence of an external magnetic field, this self-induced effective field is responsible for full paramagnetic screening, also known as the spin Meissner effect \cite{PLARubo}.
Below a certain critical value $B_c$ of the magnetic field, there is no dependence of the condensate photoluminescence energy on the magnetic field at all, and Zeeman splitting remains equal to zero.
At the same time, as the magnetic field gradually increases, it affects the circular polarization degree of the emission until the value $B_c$ is reached.
At this point, the emission becomes fully circularly polarized, and Zeeman splitting reestablishes \cite{Larionov2010,Walker2011,Fischer2014}. 

In this work, we analyze the spin Meissner effect in the system of several interacting polariton condensates. In particular, we show that the presence of TE-TM splitting in coupled spatially separated condensates results in a spin-flipping tunnel interaction and, depending on the geometry of the system, can distort the spin Meissner effect. In the simplest case of a polariton dyad, it leads to the quadratic scaling of the chemical potential with a magnetic field at small fields. In more complex triangular geometry it results in the formation of peculiar spin and polarization textures and reveal symmetry-breaking states with unequal degree of circular polarization of the nodes. Interestingly, depending on the parameters characterizing the system, cases of both positive and negative magnetic susceptibility can be realized. On the other hand, in the case of highly symmetric square configuration the spin-Meissner effect becomes restored in a broad range of system parameters. 

The paper is organized as follows. In Section~\ref{sec:2} following the Introduction, we present our model and derive the classical spin Hamiltonian. In Sections~\ref{sec:3}, \ref{sec:4}, \ref{sec:5}, we analyze the cases of the dyad, the triangle, and the square, respectively. And finally, we draw a conclusion.

\section{The Classical Hamiltonian\label{sec:2}}

We consider a system of $N$ spatially separated spinor condensates of exciton polaritons. The condensates interact with their nearest neighbors via tunneling exchange of particles, resulting in the establishment of a macroscopic phase coherence. The system is assumed to be in thermal equilibrium that favors condensation in a state with minimal energy at $T=0$. 

To calculate the energy ground state (GS) we start with the quantum Hamiltonian written in the second quantization:
\begin{eqnarray} \label{eq:Ham1}
&&\hat{H} = -\frac{\Delta_Z}{2}\sum_{j,\sigma=\pm}\sigma a^\dagger_{j\sigma}a_{j\sigma}+U\sum_{j,\sigma=\pm}a^\dagger_{j\sigma}a^\dagger_{j\sigma}a_{j\sigma}a_{j\sigma} \\
&&+J\!\!\!\sum_{\langle jl\rangle,\sigma=\pm}\!\!\!a^\dagger_{j\sigma}a_{l\sigma}+\delta J\sum_{\langle jl\rangle}\left[e^{i\theta_{lj}}a^\dagger_{j+}a_{l-}+e^{-i\theta_{lj}}a^\dagger_{l-}a_{j+}\right],\nonumber
\end{eqnarray}
where the creation (annihilation) operator $a_{j\sigma}\ (a^\dagger_{j\sigma})$  corresponds to polaritons with right and left circular polarizations (referred to as spins, $\sigma=\pm$) located at the site $j$; $\langle jl\rangle$ indicates the summation by the nearest neighbors only. 

The first term in \eqref{eq:Ham1} describes the Zeeman splitting between two opposite circular polarized components induced by an external magnetic field, while the second term corresponds to on-site polariton-polariton interactions. The intersite interaction is governed by spin-conserving tunneling with the strength $J$ and spin-flip tunneling $\delta J$ between the neighboring condensates. The latter is provided by TE-TM splitting \cite{Nalitov2015} and accounts for the difference in tunneling coefficients for polaritons linearly polarized along or transverse to the tunneling direction. In the basis of circular polarizations, this process corresponds to the tunneling with spin inversion. The angle $\theta_{jl}$ is twice the geometric angle between the $x$-axis and the line connecting the site $j$ with the site $l$ \cite{Nalitov2015}. As will be shown later, this spin-flip tunneling has a crucial impact on the condensate properties. Note that we neglected the on-site interactions of polaritons with opposite spins and interactions between the polaritons from different sites. The corresponding small terms can be easily added to our model, but do not influence the results qualitatively.

In our further consideration, we use the mean-field approach, in which the state of the coupled condensates is described by a direct product of macroscopically coherent wave functions corresponding to individual sites:
\begin{equation}
|\Psi\rangle=\prod_{j,\sigma=\pm}|\alpha_{j\sigma}\rangle,\label{ClassicalWF}
\end{equation}
where
$a_{j\sigma}|\alpha_{j\sigma}\rangle=\alpha_{j\sigma}|\alpha_{j\sigma}\rangle$ 
with order parameters being
\begin{equation}
\alpha_{j\sigma}=\sqrt{n_{j\sigma}}e^{i\varphi_{j\sigma}},
\end{equation}
where $n_{j\sigma}$ is the mean occupancy of the corresponding spin component of the $j^{\rm th}$ condensate with the phase $\varphi_{j\sigma}$. These parameters are related to total occupancies $n_j=n_{j+}+n_{j-}$ and the Stokes vector (or pseudospin) components which define the polarization of the condensate emission as follows:
\begin{subequations}
\begin{eqnarray}
    S_{jx}+iS_{jy} &=& 2\frac{\sqrt{n_{j+}n_{j-}}}{n_j}e^{i(\varphi_{j+}-\varphi_{j-})}, \\
    S_{jz} &=& \frac{n_{j+}-n_{j-}}{n_j}.
\end{eqnarray}
\end{subequations}
Note that in a coherent state, the pseudospin vector has a unit length, $S_{jx}^2+S_{jy}^2+S_{jz}^2=1$. Note that the wavefunction \eqref{ClassicalWF} represents the closest approximation of a classical state.

With these assumptions, the classical Hamiltonian of the problem reads:
\begin{equation}\label{eq.Ham0}
H(n_{j\sigma},\varphi_{j\sigma})=\langle\Psi|\hat{H}|\Psi\rangle=H_0+H_1+H_2.
\end{equation}
Here
\begin{subequations}
\begin{equation}
H_0=-\frac{\Delta_Z}{2}\sum_{j}\left(n_{j+}-n_{j-}\right)+U\sum_{j}\left(n_{j+}^2+n_{j-}^2\right)
\end{equation}
describes uncoupled condensates,
\begin{eqnarray}
    H_1=J\sum_{\langle jl\rangle}[\sqrt{n_{j+}n_{l+}}\cos(\varphi_{j+}-\varphi_{l+})+\\
    \nonumber+\sqrt{n_{j-}n_{l-}}\cos(\varphi_{j-}-\varphi_{l-})]
\end{eqnarray}
accounts for the spin-conserving tunneling, and 
\begin{eqnarray}
    H_2=\delta J\sum_{\langle jl\rangle}[\sqrt{n_{j+}n_{l-}}\cos(\varphi_{j+}-\varphi_{l-}-\theta_{jl})+\\
    \nonumber+\sqrt{n_{j-}n_{l+}}\cos(\varphi_{j-}-\varphi_{l+}+\theta_{jl})]
\end{eqnarray}
\end{subequations}
stems from the spin-flip tunneling provided by the TE-TM splitting.

Note, that as occupancy and phase are canonically conjugated variables, the dynamics of the system in a semiclassical limit is governed by a set of classical Hamilton's equations:
\begin{equation}
    \hbar\frac{d\varphi_{j\varphi}}{dt}=\frac{\partial H}{\partial n_{j\sigma}}, \qquad \hbar\frac{dn_{j\sigma}}{dt}=-\frac{\partial H}{\partial \varphi_{j\sigma}}.
\end{equation}
The analysis of the corresponding dynamics goes beyond the goals of the present paper, and in our further discussion we focus on the properties of the ground state only, where occupancies and phases reach their stationary values. 

In what follows, we consider the case of equal total occupancies, $n_{j}=n={\rm const}$. The Hamiltonian \eqref{eq.Ham0} can be then expressed in terms of phases $\varphi_{j\sigma}$ and $z$-projections of pseudospins $S_{j z}$: %It is convenient to introduce a Hamiltonian per particle $\mathcal{H}$ which gives:
\begin{equation}\label{eq.Ham}
   {H} = {H}_0+{H}_1+{H}_2
\end{equation}
with
\begin{subequations}
\begin{equation}
{H}_0=n\sum_j\left[-\frac{\Delta_Z}{2}S_{jz}+\frac{Un}{2}\left(S_{jz}^2+1\right)\right],\label{H0}    
\end{equation}
\begin{align}
    {H}_1&=  \frac{J}{2} n \sum_{\langle jl \rangle}
    \left[ 
    \sqrt{(1 + S_{jz})(1 + S_{lz})} \cos\left( \varphi_{j+}-\varphi_{l+}\right) \right. \nonumber \\
    &+  \left. \sqrt{(1 - S_{jz})(1 - S_{lz})} \cos\left( \varphi_{j-}-\varphi_{l-}\right)
    \right]\label{H1}
\end{align}
and
\begin{align}
    {H}_2 & =  \frac{\delta J}{2} n \sum_{\langle jl \rangle}
    \left[ 
    \sqrt{(1 + S_{jz})(1 - S_{lz})} \cos\left( \varphi_{j+}-\varphi_{l-}-\theta_{lj}\right) \right. \nonumber \\
    &+  \left. \sqrt{(1 - S_{jz})(1 + S_{lz})} \cos\left( \varphi_{j-}-\varphi_{l+}+\theta_{lj}\right)
    \right]. \label{H2}
\end{align}
\end{subequations}

The ground state (GS) of the system is thus defined by the minimization of the Hamiltonian \eqref{eq.Ham} in $\varphi_{j\sigma}$ and $S_{jz}$ with an additional restriction $S_{jz}\in[-1;1]$, 
\begin{equation}\label{HamiltonianMinimization}
    \frac{\partial{H}}{\partial\varphi_{j\sigma}}=0, \qquad \frac{\partial{H}}{\partial S_{jz}}=0. 
\end{equation}

The chemical potential which governs the energy of the emitted photons is then defined as:
\begin{eqnarray}\label{eq.ChemPotDef}
    \mu=&&\frac{\partial H}{\partial N}=%\frac{1}{M}\left[n\frac{\partial \mathcal{H}}{\partial n}+\mathcal{H}\right]=
    \\
   && \nonumber Un+\frac{1}{M}\left[\sum_j\left(-\frac{\Delta_Z}{2}+UnS_{jz}\right)S_{jz}+\frac{H_1}{n}+\frac{H_2}{n} \right],
\end{eqnarray}
where $N=Mn$ with $M$ being the total number of interacting condensates.

The simplest case of a single spinor condensate in a magnetic field is well studied \cite{PLARubo,Walker2011}. It features a competition between the spin-anisotropic on-site polariton-polariton interactions  which benefit from minimization of the absolute value of $S_z$, and a Zeeman splitting tending to align pseudospin along the $z$-axis. 

Minimization of $H_0$ yields that the magnetic field remains screened and does not affect the chemical potential unless the Zeeman field exceeds a critical value $\Delta_0=2Un$: 
\begin{equation}\label{eqn.SingleChemPot}
    \mu = \left\{\begin{matrix}
      & Un,  &\left|\Delta_Z\right| \leqslant \Delta_0,\\
      &  2Un - \Delta_Z/2, &\left|\Delta_Z\right| > \Delta_0.
    \end{matrix}\right.
\end{equation}

This full paramagnetic screening occurs due to the compensation of the external field by interaction-induced effective magnetic field which appears in the presence of elliptic polarization: 
\begin{equation}\label{eqn.SingleSz}
    S_z = \left\{\begin{matrix}
        {\Delta_Z}/{\Delta_0}, &\left|\Delta_Z\right| \leqslant \Delta_0, \\
        1, &\left|\Delta_Z\right|>\Delta_0.
    \end{matrix}\right. 
\end{equation}

In the following section, we demonstrate that the interaction with an adjacent condensate strongly affects the described behavior in the presence of the TE-TM energy splitting.

\section{Polariton dyad\label{sec:3}}

Let us now consider a pair of interacting condensates -- the system typically referred to as a polariton dyad. The case of a polariton dyad at zero magnetic field, corresponding to the case of an extended XY model, was studied in Ref.~\cite{kudlis2024}. Similarly to the spinless case \cite{Berloff2017}, the negative (positive) coupling $J$ results in a ferromagnetic or in-phase (antiferromagnetic with the $\pi$ phase difference) alignment in the dyad. 

In a spinor dyad, this behavior can be traced by using the total and relative phases:
\begin{align}\label{eqn.PhasesDefinition}
   \Phi_j=\frac{\varphi_{j+}+\varphi_{j-}}{2},  \quad \phi_j=\varphi_{j+}-\varphi_{j-}.
\end{align}
In this notation,  $\Phi_1 = \Phi_2$  at $J<0$ or $\Phi_1 - \Phi_2 = \pi$  at $J>0$, while the sign of a spin-flip coupling $\delta J$ defines the orientation of the condensate polarization governed by half the angle $\phi_j$. As it follows from the minimization of \eqref{H1} and \eqref{H2}, $\phi_1 = \phi_2 = \theta_{12}$  at $\delta J<0$, and the condensates are linearly polarized along the dyad edge, while their polarizations are aligned perpendicular to the edge in the opposite case of $\delta J>0$, such as $\phi_1 = \phi_2 = \theta_{12}+\pi$~\cite{kudlis2024}.

In the presence of a magnetic field, the condensates become elliptically polarized but retain identical polarization properties, $S_z = S_{1z} = S_{2z} \neq 0$, as we have checked by minimizing the total Hamiltonian \eqref{eq.Ham} numerically. The orientation of the main axis of the polarization ellipse follows the same rules as described above for $\Delta_Z=0$. 

The chemical potential reads:
\begin{multline}\label{eqn.DyadGSenergy}
       \mu=-|J|-\dfrac{\Delta_Z}{2} S_z + \dfrac{ U n}{2} (1 + S_z^2) - |\delta J|\sqrt{1 - S_z^2}.
\end{multline}
Note that the last term here plays the role of an effective in-plane  magnetic field which appears due to the linear polarization splitting from the spin-flip coupling \cite{shelykh2007SaM}. Due to this field, the pseudospin increases its in-plane component at the expense of the reduction of $S_z$. As a result, the self-interactions are unable to compensate the Zeeman field any longer, and the spin-Meissner effect is destroyed.

\begin{figure}
    \centering
    \includegraphics[width=1\linewidth]{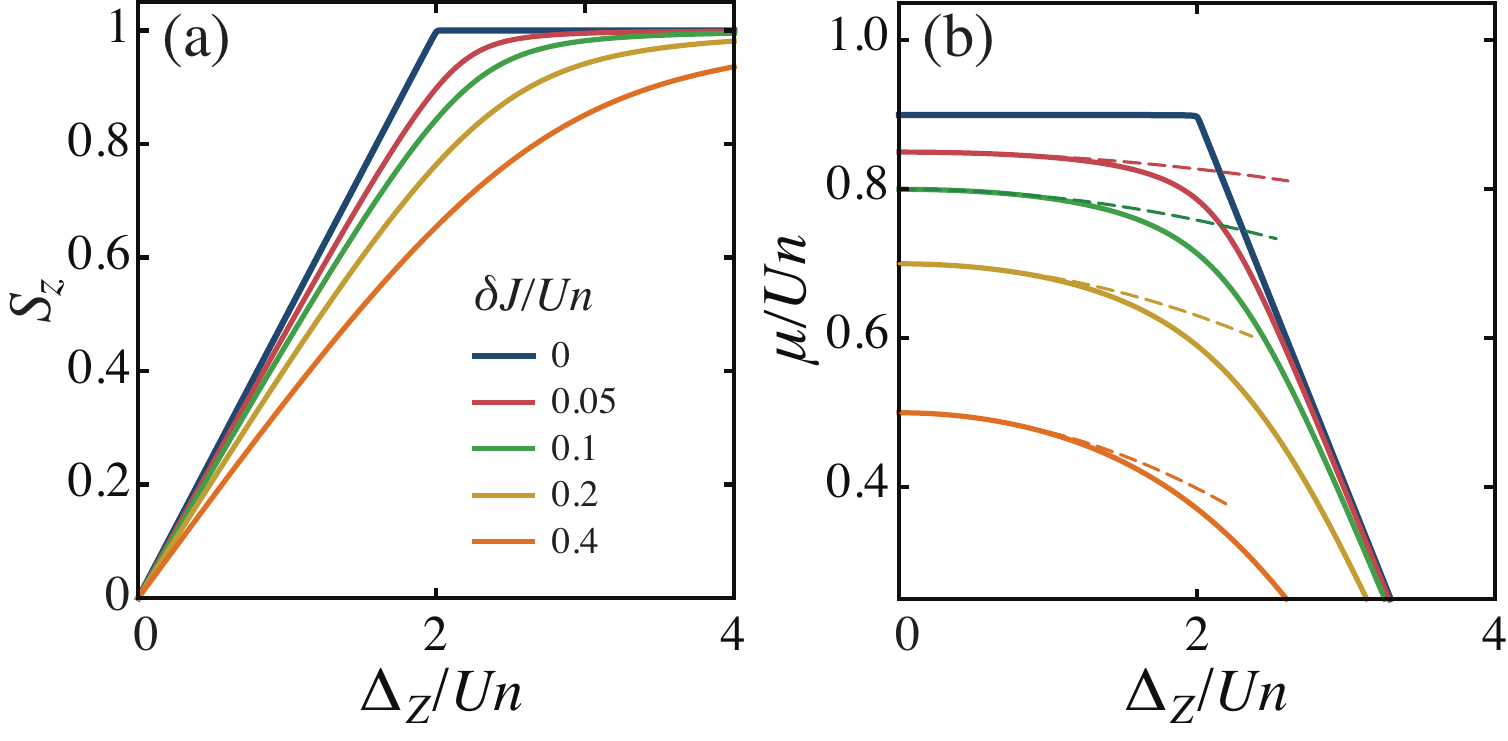}
    \caption{The circular polarization degree $S_z$ of a polariton dyad (a) and the corresponding chemical potential (b) versus bare Zeeman splitting $\Delta_Z$. The dark-blue line corresponds to the case $\delta J=0$ where  spin-dependent polariton-polariton interactions fully screen the external magnetic field and spin-Meissner effect is observed (the chemical potential does not depend on $\Delta_Z$ up to the critical value $\Delta_Z=2Un$). The colored lines illustrate the role of spin-flip tunneling, $\delta J \neq 0$. As one clearly sees, spin-Meissner effect is destroyed, so that $\mu$ is no longer independent of magnetic field at any value of $\Delta_Z$. At small fields this dependence is quadratic and is given by the approximated expression \eqref{eq.dyad.diamagnetic} (dashed lines). For big $\Delta_Z$ the dependence approaches the linear one, as in the case of a single condensate. The spin-conserving coupling is set to $J=0.1Un$.}
    \label{fig.Dyad}
\end{figure}

Naturally, at $\delta J=0$ the dependence of the chemical potential and $S_z$ on the magnetic field is the same as for the case of a single condensate, see Eqs.~\eqref{eqn.SingleChemPot} and \eqref{eqn.SingleSz} (in the expression for $\mu$ one needs simply to add a constant energy shift $-|J|$ originating from the spin-conserving coupling). The corresponding behaviour is shown in Fig.~\ref{fig.Dyad} with the thick blue line. 

However at the finite spin-flip tunneling $\delta J \neq 0$, the value of $S_z$ that minimizes the energy \eqref{eqn.DyadGSenergy} is governed by the following equation:
\begin{equation}\label{Eq.1conSz}
2 |\delta J|S_z  =  \sqrt{1 - S_z^2} (\Delta_Z - 2 U n S_z).
\end{equation}
The general solution is quite cumbersome. However, in weak Zeeman fields, one can assume that the condensates are almost linearly polarized, $|S_z| \ll 1$. In this case, Eq.~\eqref{Eq.1conSz} yields:
\begin{equation}
    S_z \approx \frac{\Delta_Z}{2\left( \left| \delta J \right| + Un \right)},
\end{equation}
and the chemical potential scales quadratically with $\Delta_Z$ for $\Delta_Z \ll U n$:
\begin{equation}\label{eq.dyad.diamagnetic}
\mu \approx Un - \left| J \right| - \left| \delta J \right| - \frac{ \left| \delta J \right|}{8\left( \left| \delta J \right| + Un \right)^2} \Delta_Z^2.
\end{equation}
In Appendix~\ref{app:sec_dyad} we also present some comments on how to obtain this solution in any order in $\Delta_Z$.

The exact values of the chemical potential $\mu$ and the circular polarization degree $S_z$ resulting from the numerical solution of Eq.~\eqref{Eq.1conSz} and the minimization of~\eqref{eqn.DyadGSenergy} are shown in Fig.~\ref{fig.Dyad}(b) together with the approximate solution for the chemical potential~\eqref{eq.dyad.diamagnetic}, see the dashed lines in Fig.~\ref{fig.Dyad}(a).

Therefore, spin-flip tunneling disrupts the paramagnetic screening effect while locking the polarization structure of the condensates. Considering the polariton dyad as a building block of polariton lattices \cite{Berloff2017}, one can naively extend this conclusion to more complex systems composed of several interacting condensates. However, in the following sections we show that this is not the case in general. 
Namely, the states with asymmetric polarization structures, as well as those reestablishing the spin-Meissner screening, can appear at the specific arrangement of the condensates.

%%%%%%%%%%%%%%%%
%%%%%%%%%%%%%%%%
\section{Triangle of polariton condensates \label{sec:4}}
%%%%%%%%%%%%%%%%
%%%%%%%%%%%%%%%%
We proceed with a more complex configuration, namely an equilateral triangle. Without loss of generality, we align the edge of the triangle with the $x$-axis such as $\theta_{12}=2\pi/3$, $\theta_{13}=0$, and $\theta_{23}=4 \pi/3$. Similarly to the case of a dyad, the case $\delta J = 0$ is equivalent to a single condensate, up to a trivial additional constant in the expression for $\mu$. 

\begin{figure*}
    \centering
    \includegraphics[width=0.75\linewidth]{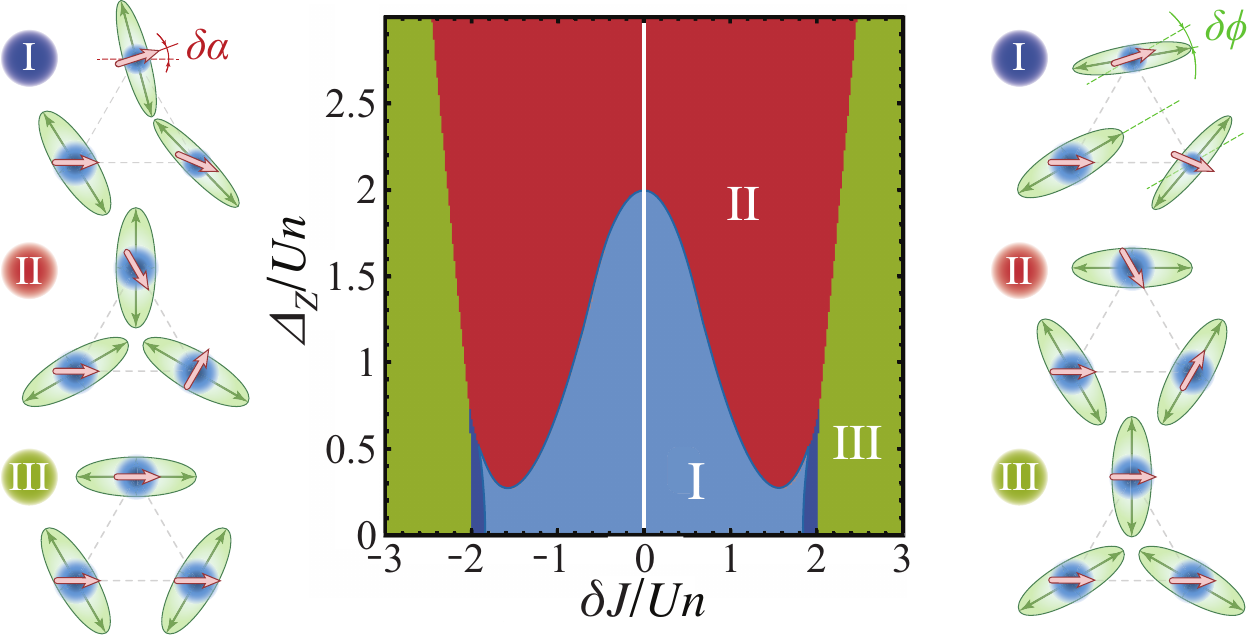}
    \caption{Phase diagram for the triangle configuration with ferromagnetic spin-conserving coupling, $J<0$. There are regions corresponding to three qualitatively different phases: the asymmetric phase I (blue region), the phase with semi-vortices II (red region), and the phase with a hidden-vortex III (green region), see main text for corresponding discussion. Note, that within the asymmetric phase there is a division into two sub-phases, which differ in sign of the magnetic susceptibility at low fields (positive in the light-blue area, which thus corresponds to a diamagnetic response, negative in the deep-blue regions). The corresponding spin and polarization configurations are shown to the right and to the left of the main panel for the cases of negative $\delta J<0$ and positive $\delta J>0$ spin-flip tunneling coefficients, respectively.  The red arrows correspond to the global phases $\Phi_j$, while green ellipses with blue spots inside illustrate elliptic polarizations of the nodes. We set $J=-Un$.}
    \label{Fig.Triangle}
\end{figure*}
 
However, at finite spin-flip coupling $\delta J \neq 0$ the situation becomes more tricky, then for a dyad. The cases of ferromagnetic ($J<0$) and antiferromagnetic ($J>0$) couplings are qualitatively different and should be considered separately. We start with the latter one, which appears to be simpler. 

For $J>0$, the ground state always corresponds to the configuration with equal circular polarization degree of all nodes, $S_{jz} = S_z$. The phases in this case are distributed as follows:
\begin{equation}\label{eqn.TriangleAFMstate}
\begin{aligned}
\varphi_{j+} &= \phi_0 - (j-1) \frac{2\pi}{3}, \\
\varphi_{j-} &= -\phi_0 + (j-1) \frac{2\pi}{3},
\end{aligned}
\end{equation}
where $j=1,2,3$ numerates the nodes. The angle $\phi_0$ that defines the orientation of the main axis of the polarization ellipses, which is arbitrary for $\delta J=0$ becomes locked at $\delta J\neq 0$ as:
\begin{equation}\label{eq:phi0PhaseIII}
     \phi_0 = \left\{\begin{matrix}
       -\pi/3,  &\delta J < 0,\\
       \pi/6, &\delta J > 0.
    \end{matrix}\right.
\end{equation}
This means that at $J>0$ the node polarizations are directed along or perpendicular to the corresponding bisectors for cases $\delta J < 0$ and $\delta J>0$ respectively.  
Note that the state \eqref{eqn.TriangleAFMstate} mimics a discrete version of the so-called hidden-vortex states \cite{Brtka2010} and has much in common with the ground state of a cavity pillar ring \cite{Sala2015} and ring-shaped polariton condensates \cite{Yulin2020,Rubo2022,Chestnov2023} in the presence of TE-TM splitting. 

The chemical potential \eqref{eqn.TriangleAFMstate} reads:
\begin{equation}\label{eqn.TriangleAFMchemPot}
   \mu=-J - \dfrac{\Delta_Z}{2}S_z + U n (1 + S_z^2) -  2|\delta J| \sqrt{1 - S_z^2},
\end{equation}
where the circular polarization degree $S_z$ obeys the condition
\begin{equation}\label{eq.TrianglePhaseIIISz}
    4 |\delta J| S_z  =  \sqrt{1 - S_z^2} (\Delta_Z - 2 U n S_z),
\end{equation}
which is analogous to Eq.~\eqref{Eq.1conSz} for the dyad. Similarly, the approximate solution for $S_z$ at the weak field reads 
\begin{equation}\label{eq.TrianglePhaseIIISzApprox}
    S_z \approx \frac{\Delta_Z}{2\left( 2 |\delta J| + Un \right)},
\end{equation}
which yields the following approximate expression for the chemical potential:
\begin{equation}\label{eq.diamagnetic}
\mu \approx Un - J - 2 |\delta J|  - \frac{ |\delta J|}{4\left(2|\delta J| + Un \right)^2} \Delta_Z^2.
\end{equation}

Unlike the case of a triangle with antiferromagnetic coupling between the nodes considered above, which is completely similar to the case of a dyad, the situation becomes much more interesting in the case of a triangle with ferromagnetic coupling ($J< 0$). The corresponding phase diagram, shown in Fig.~\ref{Fig.Triangle} reveals a set of nontrivial phases. 

The hidden vortex phase corresponding to the solution \eqref{eqn.TriangleAFMstate} still exists (green area corresponding to phase III in Fig.~\ref{Fig.Triangle}), but only for a sufficiently strong spin-flip coupling, $|\delta J |\gtrsim 2 |J|$. %In Fig.~\ref{Fig.Triangle} this phase corresponds to phase III, shown in green.

For moderate $\delta J$, two other phases appear. 

The first one is realized at weak magnetic fields corresponding to $\Delta_Z < \Delta_0$ and at $\delta J \lesssim 2J$ (phase I, blue domain).
This state evolves from the solution, which exhibits the spin Meissner effect at $\delta J = 0$, but features a structure with unequal circular polarizations of the nodes $S_{jz}$. This configuration is defined by the following conditions:
\begin{subequations}\label{eqn.TrianglePhaseIstate}
\begin{eqnarray}
(\Phi_1,\Phi_2,\Phi_3) &=& (\alpha,\alpha+\delta{\alpha},\alpha-\delta{\alpha}),\label{eqn:subst_1}\\
(\phi_1,\phi_2,\phi_3) &=& \left(2\phi_0, 2\phi_0-\delta{\phi}, 2\phi_0+\delta{\phi}\right),\label{eqn:subst_2}\\
(S_{1z},S_{2z},S_{3z})  &=& (S_z,S_z-\delta{S_z},S_z-\delta{S_z}),\label{eqn:subst_3}
\end{eqnarray}
\end{subequations}
where $\alpha$ is an arbitrary constant which can be set to 0 due to the global $U(1)$ symmetry of the problem. In this case, the angle $\phi_0$ is governed by Eq.~\eqref{eq:phi0PhaseIII} and defines the orientation of the polarization ellipse of the first condensate (the left bottom node), see Fig.~\ref{Fig.Triangle}.

\begin{figure}
    \centering
    \includegraphics[width=\linewidth]{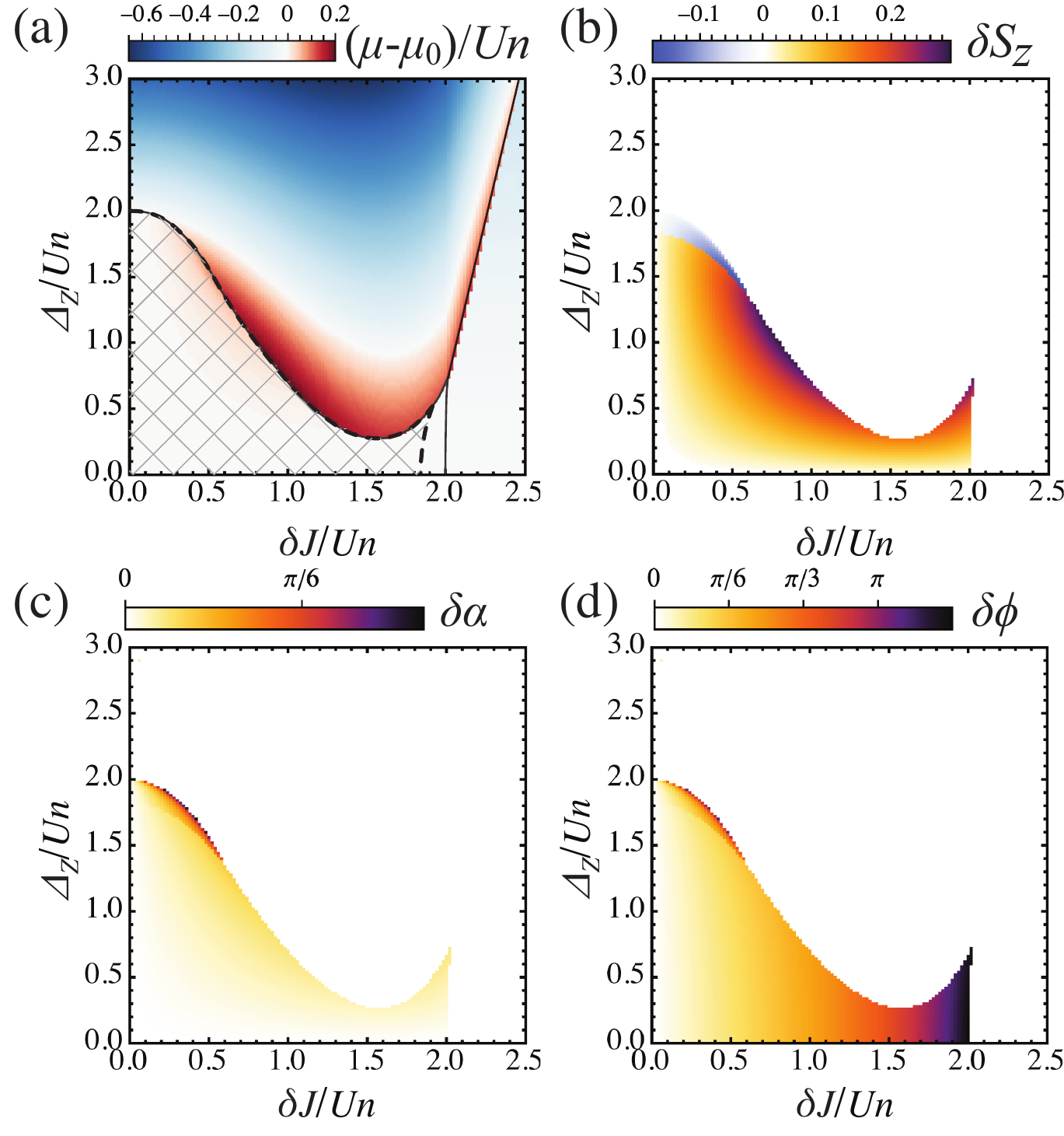}
    % \caption{Panel (a). The chemical potential $\mu$ distribution on the $\left(\delta J,\Delta_Z\right)$ phase plane. The discontinuities of the false colors highlighted with the white lines indicate phase transitions between different states. They coincide with the boundaries between the phases illustrated in Fig.~\ref{Fig.Triangle}. The region framed with the white dashed line corresponds to $\partial \mu/\partial \Delta_Z > 0$ while the opposite condition $\partial \mu/\partial \Delta_Z < 0$ holds outside it. (b) -- (d) Parameters of the state in the phase I illustrated in Fig.~\ref{Fig.Triangle}. Its structure features unequal degree of circular polarization $S_{2z}=S_{3z}\neq S_{1z}$, parameterized with the imbalance parameter $\delta S_z = S_{1z}-S_{2z}$ shown on panel (b). (c) The angle $\delta\alpha$ characterizing the difference between the total phases of the condensate: $\delta\alpha=\Phi_2 - \Phi_1 = \Phi_1-\Phi_3$ as given by Eq.~\eqref{eqn:subst_1}. (d) The double tilt angle of the condensate pseudospins $2s\delta\phi$ governed by Eq.~\eqref{eqn:subst_2}.   We set $J=-Un$. Each plot respects mirror symmetry upon changing the sign of $\delta J$. }
\caption{
(a)~Color map of the difference $\mu - \mu_0$, where $\mu$ is the chemical potential, $\mu_0$ is its value at $\Delta_Z=0$ in the $(\delta J,\Delta_Z)$ plane.
The grey solid lines highlight the phase boundaries shown in Fig.~\ref{Fig.Triangle}.
The crossed shaded area corresponds to $\partial \mu / \partial \Delta_Z > 0$.
Outside that region, we have $\partial \mu / \partial \Delta_Z < 0$. In particular, there is a small region where one has the GS in phase I and the derivative $\partial \mu / \partial \Delta_Z$ has a negative value. (b)--(d)~Key parameters of the phase~I state from Fig.~\ref{Fig.Triangle}.
In this phase, the circular polarization degrees satisfy $S_{2z} = S_{3z} \neq S_{1z}$.
We define $\delta S_z = S_{1z} - S_{2z}$ (panel~b). Panel~(c) shows $\delta \alpha$, the difference between the total phases of the condensates, given by
$\delta \alpha = \Phi_2 - \Phi_1 = \Phi_1 - \Phi_3$ [see Eq.~\eqref{eqn:subst_1}].
Panel~(d) displays the double tilt angle of the polarization ellipse $\delta \phi$ introduced in Eq.~\eqref{eqn:subst_2}, which characterizes the in-plane twist of the pseudospins. Throughout these plots, we set $J = - U n$. Each panel respects mirror symmetry upon changing the sign of $\delta J$.
}   \label{Fig.TriangleProp}
\end{figure}

The behavior of the asymmetry parameters $\delta\alpha$, $\delta\phi$ and $\delta S_z$ together with the chemical potential $\mu$ determined from the numerical analysis are shown in Fig.~\ref{Fig.TriangleProp}. To gain a better understanding of the structure of the solution discussed, we consider $\delta\alpha$, $\delta\phi$, $S_z$ and $\delta S_z$ as the variational parameters to minimize $H$. 
Then, expanding the corresponding variational equations in $\Delta_Z$ and collecting the same-order terms (see Appendix~\ref{app:sec_tr_phase_1} for details), we obtain an approximate expression for the chemical potential at small magnetic fields:
\begin{equation}\label{eqn:chem_pot_phaseI}
\mu \approx -2 |J| + U n +\dfrac{\delta J^2}{6 J} -\dfrac{\delta J^3}{36 J^2} + D(J,\delta J)\Delta_Z^2,
\end{equation}
which exhibits a quadratic field-dependence  with the corresponding coefficient being
\begin{multline}\label{eq.triad.DiamCoeff}
D(J,\delta J) = \dfrac{\delta J^2}{8 (U n)^2 (3 J- 2 U n)^2}\times\\
 \left[-3J + 4 U n -\dfrac{\delta J(J-2 U 
n)(9J^2+4 (U n)^2)}{12 J^2 (3J-2 U n)^3}\right].
\end{multline}    

\begin{figure}
    \centering
    \includegraphics[width=\linewidth]{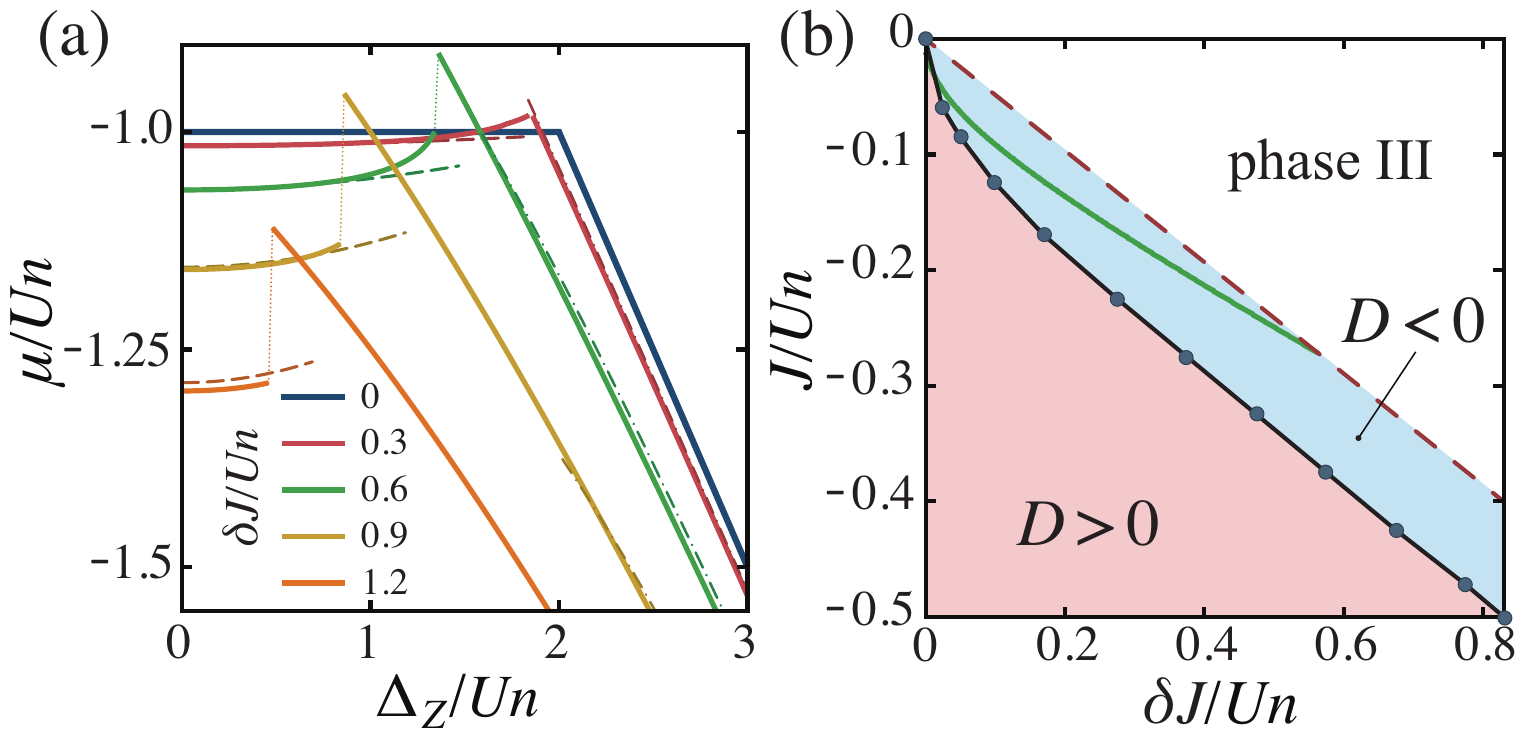}
    \caption{(a) The chemical potential of a triangle as function of the bare Zeeman splitting $\Delta_Z$ for different values of the spin-flip tunneling $\delta J$ in the regime with $D>0$. The solid curves correspond to the results of the numerical minimization procedure. The dark-blue line corresponds to the absence of spin-flip tunneling $\delta J = 0$ manifesting establishment of the spin-Meisner regime at $\Delta_Z<2Un$. The discontinuity in the curves corresponds to the abrupt switching between the phase I and phase II, see Fig.~\ref{Fig.Triangle} and Fig.~\ref{Fig.TriangleProp}(a).  The dashed curves correspond to the approximate expression \eqref{eqn:chem_pot_phaseI} in the phase I, while the dot-dashed curves correspond to Eq.~\eqref{eqn:mu_PhaseII} derived for the phase II. (b) Phase diagram for the coefficient $D=\partial\mu/\partial(\Delta_z^2)$ describing the behavior of the system in small magnetic fields. The dark-blue line with dots separating the areas of the negative (blue) and the positive (red) $D$  were extracted from the numerical simulations, while the green curve corresponds to the prediction of approximate formula~\eqref{eq.triad.DiamCoeff}. Also a dashed red line is applied, which shows where phase I is generally preserved (only in the color area). In the white area, the corresponding values of parameters ($J,\,\delta J$) already correspond to phase III.}
    \label{Fig.TriangleMu}
\end{figure}

\noindent This approximate expression demonstrates a good agreement with the exact numerical results at small and moderate values of $\delta J$. Interestingly, regimes with positive and negative $D$ can be realized, depending on the coefficients $J$ and $\delta J$ as shown in Fig.~\ref{Fig.TriangleMu}(b). 

Fig.~\ref{Fig.TriangleMu}(a) shows the $\mu(\Delta_Z)$-dependencies corresponding to the vertical line cuts (fixed $\delta J$) of the phase diagram Fig.~\ref{Fig.Triangle} for the case $D>0$ which is particularly interesting. In this regime, the chemical potential demonstrates the quadratic increase with magnetic field characteristic of the diamagnetic behavior of the system. Note that, as polaritons are electrically neutral particles, the diamagnetic-like behavior here is not related to the direct effect of the magnetic field on the orbital motion but arises from delicate interplay between Zeeman splitting, spin-dependent tunneling, and polariton-polariton interactions. The predictions of the approximated expression \eqref{eqn:chem_pot_phaseI} (the green line) agree well with the numerical results (blue dots) at moderate $\delta J$.

The discontinuities of the $\mu(\Delta_Z)$-dependencies in Fig.~\ref{Fig.TriangleMu}(a) indicate the transition to the different phase which is realized at strong enough magnetic fields. 
This state corresponds to the red region (phase II) in the phase diagram in Fig.~\ref{Fig.Triangle}. 
Unlike the previous case, it possesses a uniform spin-imbalance $S_{jz}=S_z$ similar to the solution in phase III. 

However, in contrast to the state \eqref{eqn.TriangleAFMstate}, which resembles two vortices in different spin components with opposite phase windings, it demonstrates a half-vortex structure \cite{Rubo2007PRL,Flayac2010}, so we call it semi-vortex phase. In particular, the spin component with the lowest energy ($\varphi_{j-}$ at $\Delta_Z > 0$ or $\varphi_{j+}$ at $\Delta_Z < 0$) carries the phase circulations with $2\pi/3$ phase difference between neighbors (clockwise counting). The other spin-component has no phase circulation and a zero phase difference.  In a compact form, this solution can be represented as:
\begin{eqnarray}
    &\varphi_{j\,-\sigma} &= -\sigma \phi_0, \notag\\
    &\varphi_{j\sigma} &= \sigma \left(\phi_0 + (j-1)\frac{2\pi}{3}\right),
\end{eqnarray}
where $\sigma$ gives the minus sign at $\Delta_Z >0$ and the plus sign at $\Delta_Z<0$. The angle $\phi_0$ that defines the orientation of the polarization ellipses reads:
\begin{equation}\label{eq:phi0}
     \phi_0 = \left\{\begin{matrix}
       \pi/6,  &\delta J < 0,\\
       -\pi/3, &\delta J > 0.
    \end{matrix}\right.
\end{equation}

The corresponding  arrangements of the global phases $\Phi_j$ and the polarization tilt angles $\phi_j$ are shown in Fig.~\ref{Fig.Triangle} (middle panels) for the case $\Delta_Z > 0$. Reversing the magnetic field direction results in a flip of the global phase $\Phi_j \rightarrow -\Phi_j$ only.

The chemical potential of the GS in the phase II reads: 
\begin{multline}\label{eqn:chem_pot_triangle_2phase}
   \mu=\dfrac{J}{2}\left(1+3S_z\right) - \dfrac{\Delta_Z}{2}S_z + U n (1 + S_z^2)\\
    -  |\delta J| \sqrt{1 - S_z^2},
\end{multline}
where $S_z$ is governed by:
\begin{equation}\label{Eq.2conSz}
2 |\delta J| S_z  =  \sqrt{1 - S_z^2} (\Delta_Z -3 J- 2 U n S_z).
\end{equation}
Note that unlike Eqs.~\eqref{Eq.1conSz} and \eqref{eq.TrianglePhaseIIISz} this condition involves the spin-conserving coupling strength $J$. It is convenient to search an approximate solution in the form of a series in $\delta J$ up to the second order, which gives:
\begin{align}
    S_z=1 - \dfrac{2 \delta J^2}{(|\Delta_Z| - 3 J - 2 n U)^2}.
\end{align}

This leads to the following expression for the chemical potential:
\begin{equation}
    \mu=- 2 |J| + 2U n-\dfrac{|\Delta_Z|}{2}-\dfrac{(2\sqrt{2}-1)\delta J^2}{2|\Delta_Z|}.
    \label{eqn:mu_PhaseII}
\end{equation}
This dependence exhibits a linear $\mu(\Delta_Z)$-behavior at large fields, which agrees well with our numerical findings over a large range of $\delta J$, -- see the dot-dashed lines in Fig.~\ref{Fig.TriangleMu}(a) for the analytical expression and the solid curves for the numerical results.

\section{Square of polariton condensates \label{sec:5}}

Finally, we address the configuration of four condensates placed on the vertices of a square, so that $\theta_{12}=\theta_{34} = \pi$ and $\theta_{23}=\theta_{41} = 0$. 
Since this geometry is not frustrated, there is no fundamental difference between the ferromagnetic and antiferromagnetic cases. As in the case of a dyad, inversion of the sign of the spin-conserving coupling $J$ affects the orientation of polarization but has no impact on either the symmetry of the state, its energy, or the magnetic field dependence of the chemical potential, which is what we are mainly interested in. Therefore, in what follows, we focus on the ferromagnetic case $J<0$.

\begin{figure}
    \centering
    \includegraphics[width=1\linewidth]{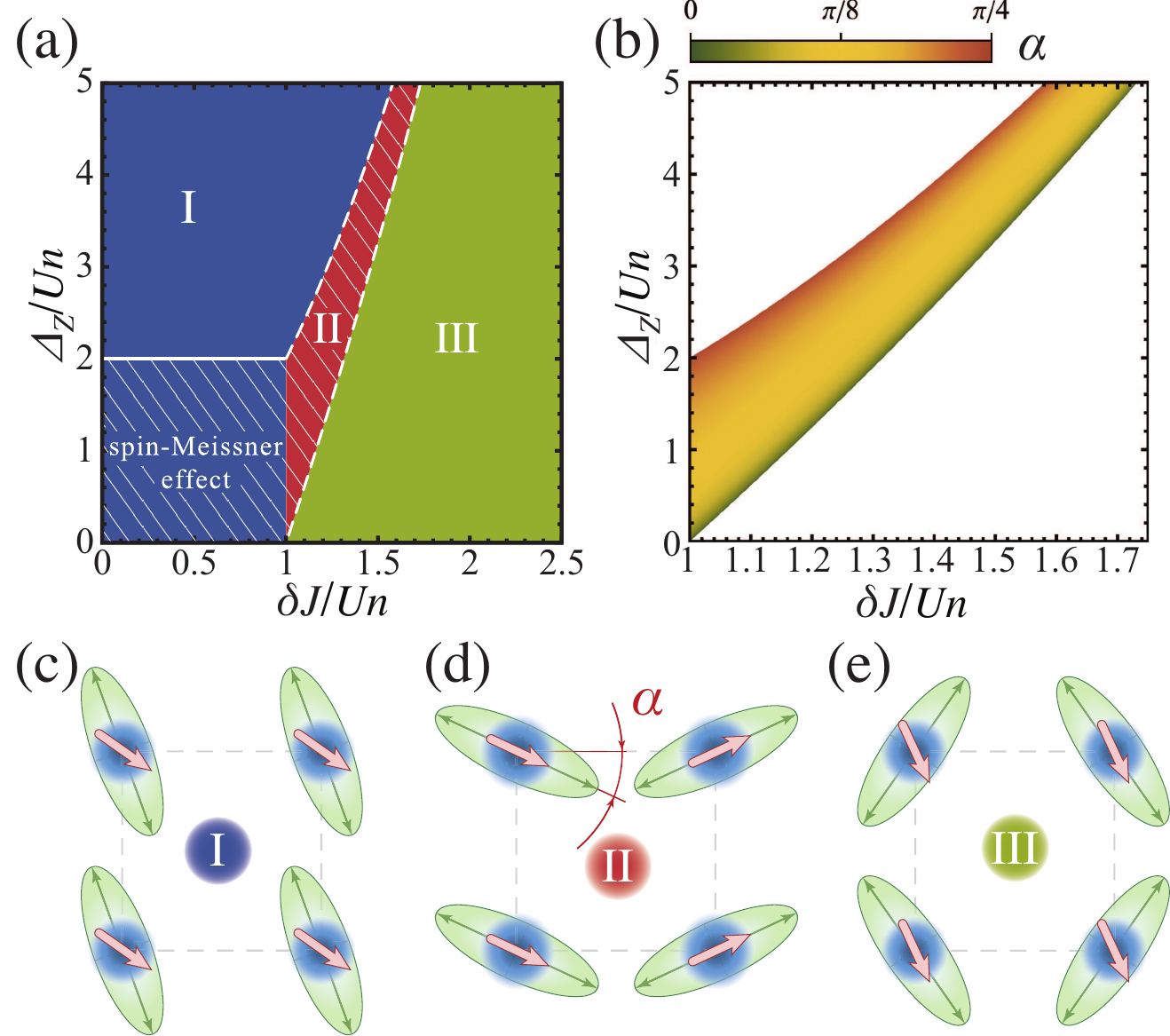}
    \caption{(a) Phase diagram for the square configuration. Blue, red and green area correspond to the uniform (I), tilted (II) and hidden vortex (III) phases, respectively. Hatching shows the domains where spin-Meissner effect is restored. We take $J=-Un$.  The boundary between red and green area is defined by Eq.~\eqref{eqn.CriticalField1}, the boundary between red and blue area  by \eqref{eqn.CriticalField2}. (b) The tilt angle $\alpha$ of the polarization ellipses $\alpha$ in phase II given by by Eq.~\eqref{eqn.SquarePhaseIIAlpha}.
    % The upper boundary corresponds to $\alpha=\pi/2$, while the lower boundary to $\alpha=\pi/4$. \AK{if we change defs: "
    The upper boundary corresponds to $\alpha=\pi/4$, while the lower boundary to $\alpha=0$.
    % " }
    (c) -- (e) Spin and polarization configurations corresponding to the different phases of the system.  The red arrows correspond to the global phases $\Phi_j$, while green ellipses with blue spots inside illustrate elliptic polarizations of the nodes.}
    \label{Fig.SquarePhaseDiagram}
\end{figure}

The most remarkable thing about square geometry is the restoration of the spin-Meissner effect for $|\delta J|<|J|$. In the corresponding phase, the states of all nodes are equivalent, 
\begin{equation}
\Phi_j=\Phi, \,\, \varphi_j=\varphi, \,\, S_{jz}=S_z, \label{SMcongiguration}   
\end{equation} 
with $\Phi, \, \varphi$ being arbitrary, and circular polarization degree defined by Eq.~\eqref{eqn.SingleSz} characteristic for a single condensate. In the phase diagram shown in Fig.~\ref{Fig.SquarePhaseDiagram} this configuration corresponds to the blue area (phase I), while the corresponding arrangement of the total phases and polaritzation ellipses is shown in Fig.~\ref{Fig.SquarePhaseDiagram}(c). 

As for the chemical potential in this phase, it is again equivalent to the case of a single condensate and is given by Eq.~\eqref{eqn.SingleChemPot}. For $\Delta_Z\leq\Delta_0$  the chemical potential is independent of the magnetic field and the spin-Meissner effect is restored (hatched area in Fig.~\ref{Fig.SquarePhaseDiagram}). 

This remarkable equivalency of a square and a single condensate for sufficiently small $\delta J$ is due to the fact that in phase \eqref{SMcongiguration} the spin-flip term rolls to zero $H_2 = 0$, since the contributions of two neighbors placed in $90 \degree$ compensate each other.
In fact, the effective magnetic field corresponding to the TE-TM splitting has a double azimuthal dependence~\cite{ShelykhReview}. As a result, changing the direction of the spin-flip tunneling by $\pi/2$ flips the sign of splitting between the states polarized along and across the tunneling direction~\cite{Nalitov2015}, and the presence of TE-TM splitting is completely screened in a square configuration, provided that all condensates are in phase.

Quite surprisingly, the spin-Meissner effect expands beyond the domain of the uniform solution (blue area). At $|\delta J| \geqslant |J|$ the uniform state \eqref{SMcongiguration} is replaced with a completely new phase (phase II, red hatched area in Fig.~\ref{Fig.SquarePhaseDiagram}) with uniform degree of circular polarization  $S_{jz} = S_z$, and phases defined as:
\begin{subequations}\label{eqn.SquarePhaseIIState}
\begin{eqnarray}
(\Phi_1,\Phi_2,\Phi_3,\Phi_4) &=& (-\alpha,-\alpha,\alpha-\pi,\alpha-\pi),\label{eqn.SquarePhaseIIState_1}\\
(\phi_1,\phi_2,\phi_3,\phi_4) &=& \left(-2\alpha,-2\alpha,2\alpha,2\alpha\right),\label{eqn.SquarePhaseIIState_2}
\end{eqnarray} 
\end{subequations}
% \begin{subequations}\label{eqn.SquarePhaseIIState}
% \begin{eqnarray}
% (\Phi_1,\Phi_2,\Phi_3,\Phi_4) &=& (\alpha,\alpha,-\alpha,-\alpha),\label{eqn.SquarePhaseIIState_1}\\
% (\phi_1,\phi_2,\phi_3,\phi_4) &=& \left(\beta,\beta,-\beta,-\beta\right), \label{eqn.SquarePhaseIIState_2}
% \end{eqnarray}
% \end{subequations}
%
where %$\beta=2\alpha\!-\!\pi$ and 
$\alpha$ governs the tilt angle of the polarization ellipses; see Fig.~\ref{Fig.SquarePhaseDiagram}(d). It is important to note that, as before, all $\Phi_j$ angles can be replaced by $\Phi_j+\alpha_0$, where $\alpha_0$ is an arbitrary angle, since $\Phi_j$ possess global U(1) rotational symmetry. Note also that the definition \eqref{eqn.SquarePhaseIIState} only stands for $\Delta_Z>0$. For the sake of brevity, here we focus on the case of positive $\Delta_Z$. The case of $\Delta_Z<0$, which demonstrates exactly the same phenomenology, has a slightly different structure, which is described in the appendix~\ref{app:sec_square}.
  
Careful analysis of Eqs.~\eqref{eqn.SquarePhaseIIState} allows us to recognize this state as two symmetric dyads with the in-phase internal arrangement. The polarization ellipses are locked and tilted by the angle $\alpha$ with respect to the line connecting the dyads, see Fig.~\ref{Fig.SquarePhaseDiagram}(d). The state \eqref{eqn.SquarePhaseIIState} exists at $|\delta J| >|J|$. With an increase of spin-flip tunneling, it merges with the hidden-vortex solution \eqref{eqn.SquareStateIII} (see below, phase III in Fig.~\ref{Fig.SquarePhaseDiagram}(a)) when each polarization ellipses are oriented along the line connecting the dyads at $\delta J>0$ or in a perpendicular direction at $\delta J < 0$, i.e. parallel to the dyad edge. Therefore, the angle $\alpha$ can be considered as an asymmetry parameter.

With the definition \eqref{eqn.SquarePhaseIIState}, the total energy \eqref{eq.Ham} reads (see Appendix~\ref{app:sec_square} for the expression valid at any sign of $\Delta_Z$):
% \begin{multline}\label{eqn:en_square_phaseII}
% E = 4n \left[-\dfrac{\Delta_Z}{2} S_z + \dfrac{U n }{2}\left(1 + S_z^2\right) \right. \\
% \left. \vphantom{\dfrac{\Delta_Z}{2}}- 2\delta J\sqrt{1 - S_z^2} \cos(2\alpha) + J \big(1 + S_z\big) \sin^2(2\alpha) \right].
% \end{multline}
% \AK{or 
\begin{multline}\label{eqn:en_square_phaseII}
E = 4n \left[-\dfrac{\Delta_Z}{2} S_z + \dfrac{U n }{2}\left(1 + S_z^2\right) \right. \\
\left. \vphantom{\dfrac{\Delta_Z}{2}}- 2\delta J\sqrt{1 - S_z^2} \cos(2\alpha) + J \big(1 + S_z\big) \sin^2(2\alpha) \right].
\end{multline}
% }
%
The corresponding variational problem with respect to $S_z$ and $\alpha$ gives:
% \begin{subequations}\label{eqn.SquarePhaseIISzandAlpha}
%     \begin{eqnarray}
%                 S_z &=&  \frac{\Delta_Z}{\Delta_0} -\frac{2\left({\delta J}^2 - J^2 \right)}{|J|\Delta_0},\label{eqn.SquarePhaseIISz} \\
%         \cos 2\alpha &=&  -\frac{\delta J}{J} \frac{\sqrt{1 - S_z^2}}{1 + S_z}.\label{eqn.SquarePhaseIIAlpha}
%     \end{eqnarray}
% \end{subequations}
% \AK{or
\begin{subequations}\label{eqn.SquarePhaseIISzandAlpha}
    \begin{eqnarray}
                S_z &=&  \frac{\Delta_Z}{\Delta_0} -\frac{2\left({\delta J}^2 - J^2 \right)}{|J|\Delta_0},\label{eqn.SquarePhaseIISz} \\
        \cos 2\alpha &=& -\frac{\delta J}{J} \frac{\sqrt{1 - S_z^2}}{1 + S_z}.\label{eqn.SquarePhaseIIAlpha}
    \end{eqnarray}
\end{subequations}
% }

\begin{figure}
    \includegraphics[width=1\linewidth]{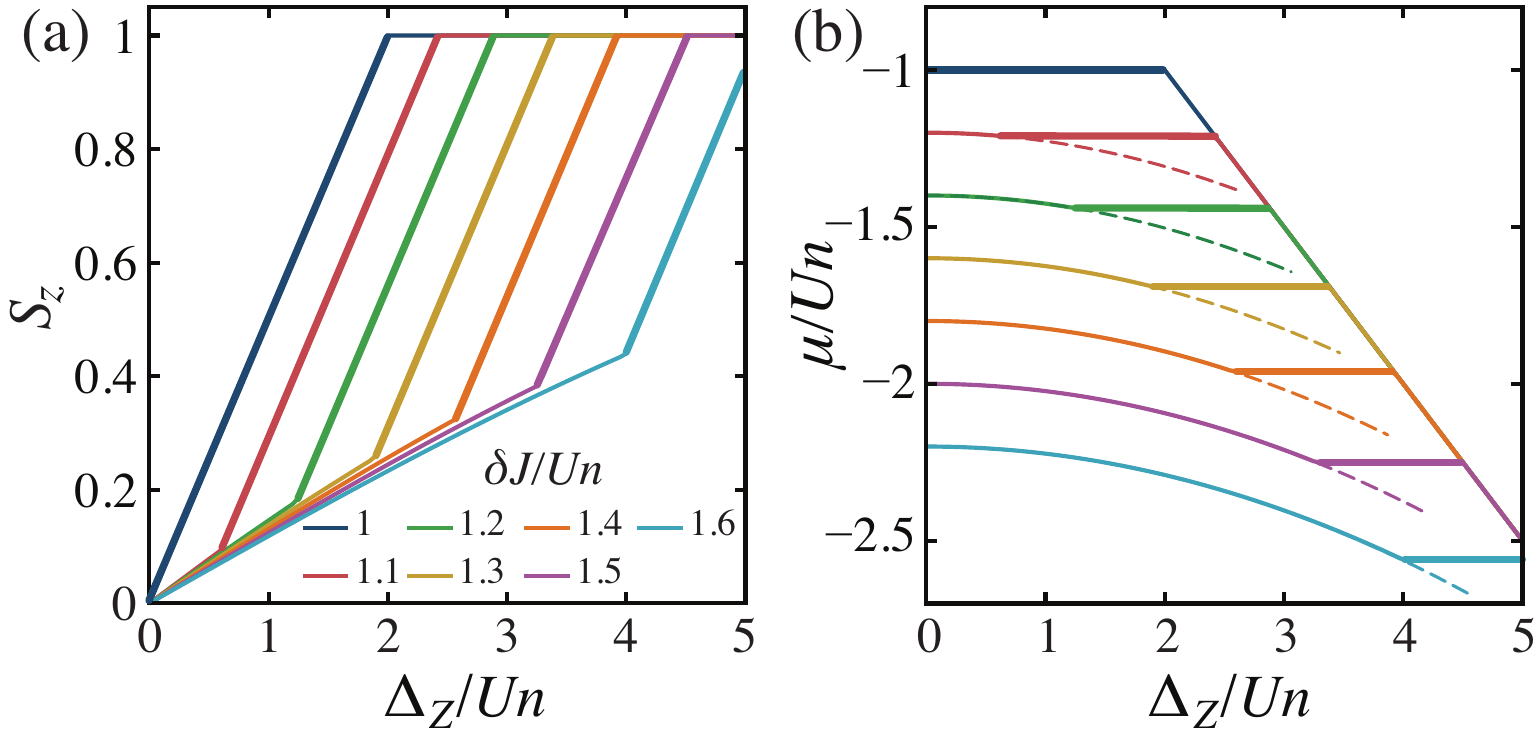}
    \caption{The degree of circular polarization $S_z$ (a) and the chemical potential (b) in the square configuration versus Zeeman field strength $\Delta_Z$ at different values of the spin-flip coupling amplitude $\delta J>|J|$. The case $\delta J \leqslant |J| = Un$ corresponds dark-blue curve. Note, that for $\delta J>|J|$ spin Meissner effect, corresponding to horizontal plateau, is restored at finite values of $\Delta_Z$. The dashed lines in panel (b) correspond to the approximated expression \eqref{eq:SquarePhaseIIIChemPot}. We set $J = - Un$.}
    \label{Fig.SquareChemPot}
\end{figure}

The expression \eqref{eqn.SquarePhaseIISz} demonstrates a linear in-field behavior of the circular polarization $S_z$, analogous to the magnetic-field screening scenario for a single condensate, see Eq.~\eqref{eqn.SingleSz}. However, due to the second term in Eq.~\eqref{eqn.SquarePhaseIISz} the spin-anisotropic interaction is unable to compensate for the magnetic-field induced energy shift. Instead, the combined effect of the spin-flip and spin-conserving coupling allows to screen the magnetic field. Indeed, substituting \eqref{eqn.SquarePhaseIISzandAlpha} into Eq.~\eqref{eqn:en_square_phaseII} results in the magnetic-field independent chemical potential:
\begin{equation}
\mu =\frac{\delta J^2}{J} + J + U n.    
\end{equation}
The corresponding behavior is shown in Fig.~\ref{Fig.SquareChemPot}(b) with the thick horizontal segments of the lines.

Finally, analogous to the case of triangle, the square configuration supports a simple uniform solution with equal phases $\Phi_j=\Phi$ and uniform spin-imbalance $S_{jz}=S_z$ in the form of the running phase state similar to state \eqref{eqn.TriangleAFMstate}:
    \begin{align}\label{eqn.SquareStateIII}
         \varphi_{j+} &= \hphantom{-} \phi_0  - (j-1) \frac{\pi}{2} ,\notag\\
        \varphi_{j-} &= - \phi_0  + (j-1) \frac{\pi}{2},
        \end{align}
%In stark contrast with the triangle configuration,    %    
where the relative angle
\begin{equation}
     \phi_0 = \left\{\begin{matrix}
       \pi/4,  &\delta J < 0,\\
       -\pi/4, &\delta J > 0
    \end{matrix}\right.
\end{equation}
corresponds to the polarization ellipses arranged along the diagonals at $\delta J > 0$ or normal to it for $\delta J <0$.
However, in stark contrast to the triangle, in a square configuration the state with opposite phase winding directions 
\begin{align}\label{eqn.SquareStateIII2}
         \varphi_{j+} &= -\phi_0  + (j-1) \frac{\pi}{2} ,\notag\\
        \varphi_{j-} &= +\phi_0  - (j-1) \frac{\pi}{2},
        \end{align}
possesses the same energy. 
%Since the states \eqref{eqn.SquareStateIII2} and \eqref{eqn.SquareStateIII} are different by the phase at each node only,
Because of this degeneracy, any linear combination of \eqref{eqn.SquareStateIII} and \eqref{eqn.SquareStateIII2} is a ground state as well. As a result, the orientation of the polarization ellipses is not fixed \cite{Chestnov2023}. Instead it obeys the following set of angles $(-\omega,\omega,-\omega,\omega)$ with an arbitrary $\omega$, see \cite{kudlis2024}.

Substituting solution \eqref{eqn.SquareStateIII} into Hamiltonian~\eqref{eq.Ham} one obtains the energy in the phase III:
\begin{align}\label{eqn:square_case3E}
    &E=4 n \left(-\dfrac{\Delta_Z}{2}S_z - 2 |\delta J| \sqrt{1 - S_z^2} + \dfrac{U n (1 + S_z^2)}{2}\right),
    % &\mu= -2 |J| +  U n, \label{eqn:square_case3mu}
\end{align}
where $S_z$ is governed by the same Eq.~\eqref{eq.TrianglePhaseIIISz} as in the case of triangle. Therefore, at the weak field, the chemical potential is given by
\begin{equation}\label{eq:SquarePhaseIIIChemPot}
\mu \approx -2 \delta J + n U - \frac{\delta J \Delta_Z^2}{4 \left(2 \delta J + n U\right)^2}.
\end{equation}
This solution is shown in Fig.~\ref{Fig.SquareChemPot}(b) with dashed lines. 

Finally, let us discuss the conditions for the transition between phases II and III. According to Fig.~\ref{Fig.SquareChemPot}(b), the state III transforms to the phase II with growth of magnetic field. The phase transition occurs when the asymmetry parameter vanishes, $\alpha=0$. From Eq.~\eqref{eqn.SquarePhaseIIAlpha} one obtains the critical magnetic field strength:
\begin{equation}\label{eqn.CriticalField1}
    \Delta_{\rm c 1} = \frac{2(J^4 - \delta J^4) + J\Delta_0 \left(\delta J^2 - J^2\right)}{J \left(\delta J^2 + J^2\right)},
\end{equation}
above which the phase II appears, -- see the bottom white dashed line in Fig.~\ref{Fig.SquarePhaseDiagram}(a). The behavior of the asymmetry parameter $\alpha$ is shown in Fig.~\ref{Fig.SquarePhaseDiagram}(b). With the increase of magnetic field, it gradually decreases to value of $\pi/4$ which corresponds to the saturation of the elliptic polarization, $S_z = 1$, see Fig.~\ref{Fig.SquareChemPot}(b) and Eqs.~\eqref{eqn.SquarePhaseIISzandAlpha}. This conditions yields the second value of the critical magnetic field:
\begin{equation}\label{eqn.CriticalField2}
    \Delta_{\rm c 2} = \Delta_0 -2\frac{\Delta J^2 - J^2}{J},
\end{equation}
at which  phase II is replaced by the phase I. This value corresponds to the upper white dashed line in the phase diagram  Fig.~\ref{Fig.SquarePhaseDiagram}(a).

\section{Conclusions}

In this work, we have studied the behavior of coupled spinor polariton condensates in a magnetic field. We have demonstrated that interplay between Zeeman splitting, spin-anisotropic polariton-polariton interactions, and different types of tunneling leads to a variety of emergent phases that have no analogs in the single-condensate case. We analyzed how the spin-Meissner effect, namely full screening of the external Zeeman splitting by spin-anisotropic polariton-polariton interaction, characteristic for a single condensate, is modified by the tunnel coupling, specifically by spin-flip terms provided by TE-TM splitting. The new polarization and phase patterns we observe (e.g., half-vortex and hidden-vortex states) can be interpreted through the interplay of the external Zeeman field with an effective in-plane field arising from TE--TM splitting, leading to the appearance of a topological phase in the wavefunction.

We demonstrated that in the simplest case of a dyad the TE-TM splitting induces an effective in-plane magnetic field which leads to the distruction of full paramegnetic screening and breakdown of spin-Meissner effect. The chemical potential in this case decreases as the square of a magnetic field at small fields, and linearly at large fields.

For more complex geometries, such as triangle and square, we have identified different polarization configurations and phase-locked sates. 

In a triangular configuration, depending on the parameters of the system, we identified the phases corresponding to the asymmetric polarization state, the half-vortex state, and the hidden-vortex state. In none of them spin-Meissner effect is present. However, surprisingly, the asymmetric polarization state demonstrates clear signatures of diamagnetic behavior: in small fields, the chemical potential increases quadratically with the field. As polaritons are electrically neutral particles, this diamagnetic-like behavior here is not related to the direct effect of the magnetic field on the orbital motion but
arises from delicate interplay between Zeeman splitting, spin-dependent tunneling, and polariton-polariton interactions.

In a highly symmetric square configuration, we demonstrated, that spin-Meissner effect is restored. For $|\delta J|< |J|$ the corresponding plateau in the chemical potential appears around $\Delta_Z=0$, while for $|\delta J|> J$ it appears at finite values of magnetic field. 

Our results provide new insights into the control of spin and phase structure in polariton lattices and highlight the potential of engineered condensate networks for spintronic and information processing applications. The ability to tune between different polarization states and restore the magnetic field screening through geometric and coupling engineering opens up promising avenues for designing robust polaritonic devices with fully controlled dynamics.

It should be noted that the obtained results are valid in the regime of partial thermal equilibrium, where the polariton condensate necessarily occupies the ground state.
In the regime of nonequilibrium bosonic condensation and, in particular, in the case of polariton lasing, the spin Meissner effect is significantly modified and can be either amplified \cite{Krol2019} or inverted, leading to alignment of the external and the effective interaction-induced magnetic fields \cite{Sawicki2024,Bochin2024}.

\section{Acknowledgements}
The work of I.Ch. (numerical analysis, interpretation of the results and writing of the text) was supported by the Russian Science Foundation grant No. 25-72-20029.
A.K. acknowledges the support of the Icelandic Research Fund (Ranns\'oknasj\'o{\dh}ur, Grant No.~2410550). A.N. acknowledges the Ministry of Science and Higher Education of the Russian Federation (Goszadaniye) Project No. FSMG-2023-0011.
We thank A.S. Melnikov for fruitful discussions. We also thank A.V. Kavokin for an important discussion concerning the interpretation of the observed effect.

\begin{appendix}

\subsection*{Appendix: Higher order corrections to chemical potential in different configurations \label{sec:app_full_system}}
\subsection{Polariton dyad}\label{app:sec_dyad}
In order to obtain the approximate expression for chemical potential at $\delta J\neq 0$, we can present the solution for $S_z$ as series in $\Delta_z$, which allows us to control the accuracy of the approximation by increasing the number of orders included. Having substituted $S_z$ as $S_z=a_0+a_1 \Delta_Z +a_2 \Delta_Z^2+a_3\Delta_Z^3+\dots$ into equation~\eqref{Eq.1conSz}, one can reexpand both sides of equation into series in $\Delta_Z$. In this case, one obtains the system of equations on coefficients $a_i$. Its solution for two lowest nonzero orders reads:
\begin{align}
    &\ a_1 = \dfrac{1}{2(|\delta J|+ n U)},\ a_3=\dfrac{-|\delta J|}{16(|\delta J| + n U)^4},
\end{align}
while $a_0$ and $a_2$ are zero. Having obtained coefficients for $S_z$ series, we can reexpand the expression for chemical potential
\begin{multline}\label{eqn:app_ferr_dyad}
    \mu=-|J| - \dfrac{\Delta_Z}{2}S_z + U n (1 + S_z^2)\\
     -  |\delta J |\sqrt{1 - S_z^2}
\end{multline}
as series in $\Delta_Z$: 
\begin{align}
    &\mu=U n -|\delta J| - |J|- \dfrac{|\delta J| \Delta_Z^2}{8 (|\delta J| + U n)^2}  \nonumber\\ 
    &\qquad\qquad\qquad\qquad\qquad +\dfrac{ (|\delta J|^2 - 3 U n|\delta J|)\Delta_Z^4}{128(|\delta J| + U n)^5},
\end{align}
where, unlike the main text, we extended the solution to the quartic order in $\Delta_Z$.

\subsection{Triangle}\label{app:sec_tr_phase_1}

Here we analyze how to obtain a solution for $\mu$ in the case of a ferromagnetic coupling in phase I of a triangular arrangement, see Fig.~\ref{Fig.Triangle}.
We should search for the analytical solution for $\delta\alpha$, $\delta\phi$, $\delta S_z$, and $S_z$ in the form of expansion in $\Delta_Z$ up to the second order ($O(\Delta_Z^3)$). After substituting these expansions into the corresponding equations (partial derivative of the Hamiltonian with respect to each of the parameters) and equating to zero the resulted expressions within each order in $\Delta_Z$ independently, we found the following structure of expansions: 
\begin{align}
&\delta{\alpha}=b_1\Delta_Z+O(\Delta_Z^3),\\
&\delta{\phi}=\omega+c_2 \Delta_Z^2+O(\Delta_Z^3),\\ 
&\delta{S_z}=a_1\Delta_Z+O(\Delta_Z^3),\\
&S_z=s_1 \Delta_Z+O(\Delta_Z^3),
\end{align}
where $b_1$, $c_2$, $a_1$, and $s_1$ are known coefficients expressed as functions of $J$ and $\delta J$. Meanwhile, for $\omega$, which is responsible for the $\Delta_Z$-independent part of $\delta{\phi}$,  one should write the separate equation which coincides with its counterpart from \cite{kudlis2024}:
\begin{equation}
\dfrac{\sqrt{3}\delta J}{J}\cos{\omega} + \left(\dfrac{\delta J}{J}+2 + 4\cos{\omega}\right)\sin{\omega}=0,\label{eq:omega_condition}
\end{equation}
the linearized solution of which (for small $\omega$) reads as follows:
\begin{align}
    \omega\approx-\dfrac{\sqrt{3} \delta J}{6 J + \delta J}.
\end{align}
Here, we do not give expressions for all coefficients due to their cumbersomeness. However, we can write out the expression for  GS energy, having first additionally expanded it for compactness in terms of the coupling $\delta J$:
\begin{multline}
E=3n\left[2J +\dfrac{\delta J^2}{6J}-\dfrac{\delta J^3}{36J^2}-\right. \\
\dfrac{1}{8 U n}
\left(1-\dfrac{\delta J^2 }{U n(3 J - 2 U n )}\right.\\
\left.\left.-\dfrac{\delta J^3\left(27 J^2-36 J U n+20 U^2 n^2 \right)}{36 J^2  U n (3 J-2  U n)^2}\right)\Delta_Z^2\right].
\end{multline}
Its derivative \eqref{eq.ChemPotDef} defines a chemical potential given by Eq.~\eqref{eqn:chem_pot_phaseI} in the main text.

\subsection{Square}\label{app:sec_square}
Here we give an extended description of the non-trivial state in the phase II of a square arrangement, which supports spin-Meissner effect above the critical field, $\Delta_Z > \Delta_0$, see Fig.~\ref{Fig.SquarePhaseDiagram}(a). At $\Delta_Z>0$, the symmetry of the phase and polarization configuration can be captured by the following conditions:
\begin{subequations}\label{eqn.SquarePhaseIIState0}
\begin{eqnarray}
(\Phi_1,\Phi_2,\Phi_3,\Phi_4) &=& (-\alpha,-\alpha,\alpha-\pi,\alpha-\pi),\label{eqn.SquarePhaseIIState_1}\\
(\phi_1,\phi_2,\phi_3,\phi_4) &=& \left(-2\alpha,-2\alpha,2\alpha,2\alpha\right),\label{eqn.SquarePhaseIIState_2}
%(S_{1z},S_{2z},S_{3z},S_{4z})  &=& (S_z,S_z,S_z,S_z),\label{eqn.SquarePhaseIIState_3}
% &(\xi_1,\xi_2,\xi_3)=(\xi,\xi+\delta_{\xi},\xi+\delta_{\xi}),\label{eqn:subst_3}
\end{eqnarray}
\end{subequations}
while for $\Delta_Z<0$ it reads 
\begin{subequations}\label{eqn.SquarePhaseIIState_2}
\begin{eqnarray}
(\Phi_1,\Phi_2,\Phi_3,\Phi_4) &=& (-\alpha,-\alpha,\pi-3\alpha,\pi-3\alpha),\label{eqn.SquarePhaseIIState_2_1}\\
(\phi_1,\phi_2,\phi_3,\phi_4) &=& \left(-2\alpha,-2\alpha,2\alpha,2\alpha\right).\label{eqn.SquarePhaseIIState_2_2}
\end{eqnarray}
\end{subequations}
In both cases all condensates have an identical  degree of circular polarization, $S_{jz} = S_z$.

With this parametrization, the energy acquires the following form:
\begin{multline}\label{eqn:en_square_phase2}
E = 4n \bigg[-\dfrac{\Delta_Z}{2} S_z + \dfrac{U n }{2}\left(1 + S_z^2\right) \\
- 2\delta J\sqrt{1 - S_z^2} \cos(2\alpha) + J \big(1 + |S_z|\big) \sin^2(2\alpha) \bigg],
\end{multline}
while the chemical potential reads
\begin{multline}
\mu = -\frac{\Delta_Z }{2}S_z + Un \left(1 + S_z^2\right)  \\
- 2\delta J \sqrt{1 - S_z^2} \cos(2\alpha) + J \big(1 + |S_z|\big) \sin^2(2\alpha).
\end{multline}
The $\alpha$ angle and $S_z$ can be found from minimization of~\eqref{eqn:en_square_phase2}. It turns out that the corresponding system has an exact analytical solution with minimal energy:
\begin{subequations}
    \begin{eqnarray}
        S_z &=& {\rm sign}(\Delta_Z)\left[\frac{|\Delta_Z|}{\Delta_0} + \frac{2\left({\delta J}^2 - J^2 \right)}{J\Delta_0}\right], \\
        \cos 2\alpha &=&  -\frac{\delta J}{J} \frac{\sqrt{1 - S_z^2}}{1 + |S_z|}.
    \end{eqnarray}
\end{subequations}
which gives the following expression for the energy and chemical potential:
\begin{align}
E&=4n\!\! \left[\!\frac{\delta J^2}{J} + J - \frac{\left(2\delta J^2 + (|\Delta_Z| - 2J)J\right)^2}{8J^2 U n} + \frac{Un}{2} \!\right]\!\!,\!\!\\
\mu&=
\frac{\delta J^2}{J} + J + U n.
\end{align}

\end{appendix}

%\bibliography{references}   

\begin{thebibliography}{63}%
\makeatletter
\providecommand \@ifxundefined [1]{%
 \@ifx{#1\undefined}
}%
\providecommand \@ifnum [1]{%
 \ifnum #1\expandafter \@firstoftwo
 \else \expandafter \@secondoftwo
 \fi
}%
\providecommand \@ifx [1]{%
 \ifx #1\expandafter \@firstoftwo
 \else \expandafter \@secondoftwo
 \fi
}%
\providecommand \natexlab [1]{#1}%
\providecommand \enquote  [1]{``#1''}%
\providecommand \bibnamefont  [1]{#1}%
\providecommand \bibfnamefont [1]{#1}%
\providecommand \citenamefont [1]{#1}%
\providecommand \href@noop [0]{\@secondoftwo}%
\providecommand \href [0]{\begingroup \@sanitize@url \@href}%
\providecommand \@href[1]{\@@startlink{#1}\@@href}%
\providecommand \@@href[1]{\endgroup#1\@@endlink}%
\providecommand \@sanitize@url [0]{\catcode `\\12\catcode `\$12\catcode
  `\&12\catcode `\#12\catcode `\^12\catcode `\_12\catcode `\%12\relax}%
\providecommand \@@startlink[1]{}%
\providecommand \@@endlink[0]{}%
\providecommand \url  [0]{\begingroup\@sanitize@url \@url }%
\providecommand \@url [1]{\endgroup\@href {#1}{\urlprefix }}%
\providecommand \urlprefix  [0]{URL }%
\providecommand \Eprint [0]{\href }%
\providecommand \doibase [0]{https://doi.org/}%
\providecommand \selectlanguage [0]{\@gobble}%
\providecommand \bibinfo  [0]{\@secondoftwo}%
\providecommand \bibfield  [0]{\@secondoftwo}%
\providecommand \translation [1]{[#1]}%
\providecommand \BibitemOpen [0]{}%
\providecommand \bibitemStop [0]{}%
\providecommand \bibitemNoStop [0]{.\EOS\space}%
\providecommand \EOS [0]{\spacefactor3000\relax}%
\providecommand \BibitemShut  [1]{\csname bibitem#1\endcsname}%
\let\auto@bib@innerbib\@empty
%</preamble>
\bibitem [{\citenamefont {Balents}(2010)}]{Balents2010}%
  \BibitemOpen
  \bibfield  {author} {\bibinfo {author} {\bibfnamefont {L.}~\bibnamefont
  {Balents}},\ }\bibfield  {title} {\bibinfo {title} {Spin liquids in
  frustrated magnets},\ }\href {https://doi.org/10.1038/nature08917} {\bibfield
   {journal} {\bibinfo  {journal} {Nature}\ }\textbf {\bibinfo {volume}
  {464}},\ \bibinfo {pages} {199} (\bibinfo {year} {2010})}\BibitemShut
  {NoStop}%
\bibitem [{\citenamefont {Lee}\ \emph {et~al.}(2006)\citenamefont {Lee},
  \citenamefont {Nagaosa},\ and\ \citenamefont {Wen}}]{PatrickLee2006}%
  \BibitemOpen
  \bibfield  {author} {\bibinfo {author} {\bibfnamefont {P.~A.}\ \bibnamefont
  {Lee}}, \bibinfo {author} {\bibfnamefont {N.}~\bibnamefont {Nagaosa}},\ and\
  \bibinfo {author} {\bibfnamefont {X.-G.}\ \bibnamefont {Wen}},\ }\bibfield
  {title} {\bibinfo {title} {Doping a {M}ott insulator: Physics of
  high-temperature superconductivity},\ }\href
  {https://doi.org/10.1103/RevModPhys.78.17} {\bibfield  {journal} {\bibinfo
  {journal} {Rev. Mod. Phys.}\ }\textbf {\bibinfo {volume} {78}},\ \bibinfo
  {pages} {17} (\bibinfo {year} {2006})}\BibitemShut {NoStop}%
\bibitem [{\citenamefont {Das}\ and\ \citenamefont {Balatsky}(2013)}]{Das2013}%
  \BibitemOpen
  \bibfield  {author} {\bibinfo {author} {\bibfnamefont {T.}~\bibnamefont
  {Das}}\ and\ \bibinfo {author} {\bibfnamefont {A.~V.}\ \bibnamefont
  {Balatsky}},\ }\bibfield  {title} {\bibinfo {title} {Engineering
  three-dimensional topological insulators in {R}ashba-type spin-orbit coupled
  heterostructures},\ }\href {https://doi.org/10.1038/ncomms2972} {\bibfield
  {journal} {\bibinfo  {journal} {Nat. Commun.}\ }\textbf {\bibinfo {volume}
  {4}},\ \bibinfo {pages} {1972} (\bibinfo {year} {2013})}\BibitemShut
  {NoStop}%
\bibitem [{\citenamefont {Bloch}\ \emph {et~al.}(2008)\citenamefont {Bloch},
  \citenamefont {Dalibard},\ and\ \citenamefont {Zwerger}}]{Bloch2008}%
  \BibitemOpen
  \bibfield  {author} {\bibinfo {author} {\bibfnamefont {I.}~\bibnamefont
  {Bloch}}, \bibinfo {author} {\bibfnamefont {J.}~\bibnamefont {Dalibard}},\
  and\ \bibinfo {author} {\bibfnamefont {W.}~\bibnamefont {Zwerger}},\
  }\bibfield  {title} {\bibinfo {title} {Many-body physics with ultracold
  gases},\ }\href {https://doi.org/10.1103/RevModPhys.80.885} {\bibfield
  {journal} {\bibinfo  {journal} {Rev. Mod. Phys.}\ }\textbf {\bibinfo {volume}
  {80}},\ \bibinfo {pages} {885} (\bibinfo {year} {2008})}\BibitemShut
  {NoStop}%
\bibitem [{\citenamefont {Greiner}\ \emph {et~al.}(2002)\citenamefont
  {Greiner}, \citenamefont {Mandel}, \citenamefont {Esslinger}, \citenamefont
  {H\"ansch},\ and\ \citenamefont {Bloch}}]{Greiner2002}%
  \BibitemOpen
  \bibfield  {author} {\bibinfo {author} {\bibfnamefont {M.}~\bibnamefont
  {Greiner}}, \bibinfo {author} {\bibfnamefont {M.~O.}\ \bibnamefont {Mandel}},
  \bibinfo {author} {\bibfnamefont {T.}~\bibnamefont {Esslinger}}, \bibinfo
  {author} {\bibnamefont {H\"ansch}},\ and\ \bibinfo {author} {\bibfnamefont
  {I.}~\bibnamefont {Bloch}},\ }\bibfield  {title} {\bibinfo {title} {Quantum
  phase transition from a superfluid to a {M}ott insulator in a gas of
  ultracold atoms},\ }\href {https://doi.org/10.1038/415039a} {\bibfield
  {journal} {\bibinfo  {journal} {Nature}\ }\textbf {\bibinfo {volume} {415}},\
  \bibinfo {pages} {39} (\bibinfo {year} {2002})}\BibitemShut {NoStop}%
\bibitem [{\citenamefont {Kinoshita}\ \emph {et~al.}(2004)\citenamefont
  {Kinoshita}, \citenamefont {Wenger},\ and\ \citenamefont
  {Weiss}}]{Kinoshita2004}%
  \BibitemOpen
  \bibfield  {author} {\bibinfo {author} {\bibfnamefont {T.}~\bibnamefont
  {Kinoshita}}, \bibinfo {author} {\bibfnamefont {T.}~\bibnamefont {Wenger}},\
  and\ \bibinfo {author} {\bibfnamefont {D.~S.}\ \bibnamefont {Weiss}},\
  }\bibfield  {title} {\bibinfo {title} {Observation of a one-dimensional
  {T}onks-{G}irardeau gas},\ }\href {https://doi.org/10.1126/science.1100700}
  {\bibfield  {journal} {\bibinfo  {journal} {Science}\ }\textbf {\bibinfo
  {volume} {305}},\ \bibinfo {pages} {1125} (\bibinfo {year}
  {2004})}\BibitemShut {NoStop}%
\bibitem [{\citenamefont {Paredes}\ \emph {et~al.}(2004)\citenamefont
  {Paredes}, \citenamefont {Widera}, \citenamefont {Murg}, \citenamefont
  {F\"olling}, \citenamefont {Cirac}, \citenamefont {Shlyapnikov},
  \citenamefont {H\"ansch},\ and\ \citenamefont {Bloch}}]{Paredes2004}%
  \BibitemOpen
  \bibfield  {author} {\bibinfo {author} {\bibfnamefont {B.}~\bibnamefont
  {Paredes}}, \bibinfo {author} {\bibfnamefont {A.}~\bibnamefont {Widera}},
  \bibinfo {author} {\bibfnamefont {V.}~\bibnamefont {Murg}}, \bibinfo {author}
  {\bibfnamefont {S.}~\bibnamefont {F\"olling}}, \bibinfo {author}
  {\bibfnamefont {J.~I.}\ \bibnamefont {Cirac}}, \bibinfo {author}
  {\bibfnamefont {G.~V.}\ \bibnamefont {Shlyapnikov}}, \bibinfo {author}
  {\bibfnamefont {T.~W.}\ \bibnamefont {H\"ansch}},\ and\ \bibinfo {author}
  {\bibfnamefont {I.}~\bibnamefont {Bloch}},\ }\bibfield  {title} {\bibinfo
  {title} {{T}onks–{G}irardeau gas of ultracold atoms in an optical
  lattice},\ }\href {https://doi.org/10.1038/nature02530} {\bibfield  {journal}
  {\bibinfo  {journal} {Nature}\ }\textbf {\bibinfo {volume} {429}},\ \bibinfo
  {pages} {277} (\bibinfo {year} {2004})}\BibitemShut {NoStop}%
\bibitem [{\citenamefont {Morsch}\ \emph {et~al.}(2001)\citenamefont {Morsch},
  \citenamefont {M\"uller}, \citenamefont {Cristiani}, \citenamefont
  {Ciampini},\ and\ \citenamefont {Arimondo}}]{Morsch2001}%
  \BibitemOpen
  \bibfield  {author} {\bibinfo {author} {\bibfnamefont {O.}~\bibnamefont
  {Morsch}}, \bibinfo {author} {\bibfnamefont {J.~H.}\ \bibnamefont
  {M\"uller}}, \bibinfo {author} {\bibfnamefont {M.}~\bibnamefont {Cristiani}},
  \bibinfo {author} {\bibfnamefont {D.}~\bibnamefont {Ciampini}},\ and\
  \bibinfo {author} {\bibfnamefont {E.}~\bibnamefont {Arimondo}},\ }\bibfield
  {title} {\bibinfo {title} {{B}loch oscillations and mean-field effects of
  {B}ose-{E}instein condensates in {1D} optical lattices},\ }\href
  {https://doi.org/10.1103/PhysRevLett.87.140402} {\bibfield  {journal}
  {\bibinfo  {journal} {Phys. Rev. Lett.}\ }\textbf {\bibinfo {volume} {87}},\
  \bibinfo {pages} {140402} (\bibinfo {year} {2001})}\BibitemShut {NoStop}%
\bibitem [{\citenamefont {Simon}\ \emph {et~al.}(2004)\citenamefont {Simon},
  \citenamefont {Bakr}, \citenamefont {Ma}, \citenamefont {Tai}, \citenamefont
  {Preiss},\ and\ \citenamefont {Greiner}}]{Simon2011}%
  \BibitemOpen
  \bibfield  {author} {\bibinfo {author} {\bibfnamefont {J.}~\bibnamefont
  {Simon}}, \bibinfo {author} {\bibfnamefont {W.~S.}\ \bibnamefont {Bakr}},
  \bibinfo {author} {\bibfnamefont {R.}~\bibnamefont {Ma}}, \bibinfo {author}
  {\bibfnamefont {M.~E.}\ \bibnamefont {Tai}}, \bibinfo {author} {\bibfnamefont
  {P.~M.}\ \bibnamefont {Preiss}},\ and\ \bibinfo {author} {\bibfnamefont
  {M.}~\bibnamefont {Greiner}},\ }\bibfield  {title} {\bibinfo {title} {Quantum
  simulation of antiferromagnetic spin chains in an optical lattice},\ }\href
  {https://doi.org/10.1038/nature09994} {\bibfield  {journal} {\bibinfo
  {journal} {Nature}\ }\textbf {\bibinfo {volume} {472}},\ \bibinfo {pages}
  {307} (\bibinfo {year} {2004})}\BibitemShut {NoStop}%
\bibitem [{\citenamefont {Trompeter}\ \emph {et~al.}(2006)\citenamefont
  {Trompeter}, \citenamefont {Krolikowski}, \citenamefont {Neshev},
  \citenamefont {Desyatnikov}, \citenamefont {Sukhorukov}, \citenamefont
  {Kivshar}, \citenamefont {Pertsch}, \citenamefont {Peschel},\ and\
  \citenamefont {Lederer}}]{Trompeter2006}%
  \BibitemOpen
  \bibfield  {author} {\bibinfo {author} {\bibfnamefont {H.}~\bibnamefont
  {Trompeter}}, \bibinfo {author} {\bibfnamefont {W.}~\bibnamefont
  {Krolikowski}}, \bibinfo {author} {\bibfnamefont {D.~N.}\ \bibnamefont
  {Neshev}}, \bibinfo {author} {\bibfnamefont {A.~S.}\ \bibnamefont
  {Desyatnikov}}, \bibinfo {author} {\bibfnamefont {A.~A.}\ \bibnamefont
  {Sukhorukov}}, \bibinfo {author} {\bibfnamefont {Y.~S.}\ \bibnamefont
  {Kivshar}}, \bibinfo {author} {\bibfnamefont {T.}~\bibnamefont {Pertsch}},
  \bibinfo {author} {\bibfnamefont {U.}~\bibnamefont {Peschel}},\ and\ \bibinfo
  {author} {\bibfnamefont {F.}~\bibnamefont {Lederer}},\ }\bibfield  {title}
  {\bibinfo {title} {{B}loch oscillations and {Z}ener tunneling in
  two-dimensional photonic lattices},\ }\href
  {https://doi.org/10.1103/PhysRevLett.96.053903} {\bibfield  {journal}
  {\bibinfo  {journal} {Phys. Rev. Lett.}\ }\textbf {\bibinfo {volume} {96}},\
  \bibinfo {pages} {053903} (\bibinfo {year} {2006})}\BibitemShut {NoStop}%
\bibitem [{\citenamefont {Khanikaev}\ \emph {et~al.}(2013)\citenamefont
  {Khanikaev}, \citenamefont {Mousavi}, \citenamefont {Wang-Kong},
  \citenamefont {Kargarian}, \citenamefont {MacDonald},\ and\ \citenamefont
  {Shvets}}]{Khanikaev2013}%
  \BibitemOpen
  \bibfield  {author} {\bibinfo {author} {\bibfnamefont {A.~B.}\ \bibnamefont
  {Khanikaev}}, \bibinfo {author} {\bibfnamefont {S.~H.}\ \bibnamefont
  {Mousavi}}, \bibinfo {author} {\bibfnamefont {T.}~\bibnamefont {Wang-Kong}},
  \bibinfo {author} {\bibfnamefont {M.}~\bibnamefont {Kargarian}}, \bibinfo
  {author} {\bibfnamefont {A.~H.}\ \bibnamefont {MacDonald}},\ and\ \bibinfo
  {author} {\bibfnamefont {G.}~\bibnamefont {Shvets}},\ }\bibfield  {title}
  {\bibinfo {title} {Photonic topological insulators},\ }\href
  {https://doi.org/10.1038/nmat3520} {\bibfield  {journal} {\bibinfo  {journal}
  {Nat. Mater.}\ }\textbf {\bibinfo {volume} {12}},\ \bibinfo {pages} {233}
  (\bibinfo {year} {2013})}\BibitemShut {NoStop}%
\bibitem [{\citenamefont {Rechtsman}\ \emph {et~al.}(2013)\citenamefont
  {Rechtsman}, \citenamefont {Zeuner}, \citenamefont {Plotnik}, \citenamefont
  {Lumer}, \citenamefont {Podolsky}, \citenamefont {Dreisow}, \citenamefont
  {Nolte}, \citenamefont {Segev},\ and\ \citenamefont
  {Szameit}}]{Rechtsman2013}%
  \BibitemOpen
  \bibfield  {author} {\bibinfo {author} {\bibfnamefont {M.~C.}\ \bibnamefont
  {Rechtsman}}, \bibinfo {author} {\bibfnamefont {J.~M.}\ \bibnamefont
  {Zeuner}}, \bibinfo {author} {\bibfnamefont {Y.}~\bibnamefont {Plotnik}},
  \bibinfo {author} {\bibfnamefont {Y.}~\bibnamefont {Lumer}}, \bibinfo
  {author} {\bibfnamefont {D.}~\bibnamefont {Podolsky}}, \bibinfo {author}
  {\bibfnamefont {F.}~\bibnamefont {Dreisow}}, \bibinfo {author} {\bibfnamefont
  {S.}~\bibnamefont {Nolte}}, \bibinfo {author} {\bibfnamefont
  {M.}~\bibnamefont {Segev}},\ and\ \bibinfo {author} {\bibfnamefont
  {A.}~\bibnamefont {Szameit}},\ }\bibfield  {title} {\bibinfo {title}
  {Photonic {F}loquet topological insulators},\ }\href
  {https://doi.org/10.1038/nature12066} {\bibfield  {journal} {\bibinfo
  {journal} {Nature}\ }\textbf {\bibinfo {volume} {496}},\ \bibinfo {pages}
  {196} (\bibinfo {year} {2013})}\BibitemShut {NoStop}%
\bibitem [{\citenamefont {Liang}\ and\ \citenamefont
  {Chong}(2013)}]{Liang2013}%
  \BibitemOpen
  \bibfield  {author} {\bibinfo {author} {\bibfnamefont {G.~Q.}\ \bibnamefont
  {Liang}}\ and\ \bibinfo {author} {\bibfnamefont {Y.~D.}\ \bibnamefont
  {Chong}},\ }\bibfield  {title} {\bibinfo {title} {Optical resonator analog of
  a two-dimensional topological insulator},\ }\href
  {https://doi.org/10.1103/PhysRevLett.110.203904} {\bibfield  {journal}
  {\bibinfo  {journal} {Phys. Rev. Lett.}\ }\textbf {\bibinfo {volume} {110}},\
  \bibinfo {pages} {203904} (\bibinfo {year} {2013})}\BibitemShut {NoStop}%
\bibitem [{\citenamefont {Fleischer}\ \emph {et~al.}(2003)\citenamefont
  {Fleischer}, \citenamefont {Segev}, \citenamefont {Efremidis},\ and\
  \citenamefont {Christodoulides}}]{Fleischer2003}%
  \BibitemOpen
  \bibfield  {author} {\bibinfo {author} {\bibfnamefont {J.~W.}\ \bibnamefont
  {Fleischer}}, \bibinfo {author} {\bibfnamefont {M.}~\bibnamefont {Segev}},
  \bibinfo {author} {\bibfnamefont {N.~K.}\ \bibnamefont {Efremidis}},\ and\
  \bibinfo {author} {\bibfnamefont {D.~N.}\ \bibnamefont {Christodoulides}},\
  }\bibfield  {title} {\bibinfo {title} {Observation of two-dimensional
  discrete solitons in optically induced nonlinear photonic lattices},\ }\href
  {https://doi.org/10.1038/nature01452} {\bibfield  {journal} {\bibinfo
  {journal} {Nature}\ }\textbf {\bibinfo {volume} {422}},\ \bibinfo {pages}
  {147} (\bibinfo {year} {2003})}\BibitemShut {NoStop}%
\bibitem [{\citenamefont {Li}\ \emph {et~al.}(2023)\citenamefont {Li},
  \citenamefont {Koniakhin}, \citenamefont {Nalitov}, \citenamefont
  {Cherotchenko}, \citenamefont {Solnyshkov}, \citenamefont {Malpuech},
  \citenamefont {Xiao}, \citenamefont {Zhang},\ and\ \citenamefont
  {Zhang}}]{Li2023}%
  \BibitemOpen
  \bibfield  {author} {\bibinfo {author} {\bibfnamefont {F.}~\bibnamefont
  {Li}}, \bibinfo {author} {\bibfnamefont {S.~V.}\ \bibnamefont {Koniakhin}},
  \bibinfo {author} {\bibfnamefont {A.~V.}\ \bibnamefont {Nalitov}}, \bibinfo
  {author} {\bibfnamefont {E.}~\bibnamefont {Cherotchenko}}, \bibinfo {author}
  {\bibfnamefont {D.~D.}\ \bibnamefont {Solnyshkov}}, \bibinfo {author}
  {\bibfnamefont {G.}~\bibnamefont {Malpuech}}, \bibinfo {author}
  {\bibfnamefont {M.}~\bibnamefont {Xiao}}, \bibinfo {author} {\bibfnamefont
  {Y.}~\bibnamefont {Zhang}},\ and\ \bibinfo {author} {\bibfnamefont
  {Z.}~\bibnamefont {Zhang}},\ }\bibfield  {title} {\bibinfo {title}
  {Simultaneous creation of multiple vortex-antivortex pairs in momentum space
  in photonic lattices},\ }\bibfield  {journal} {\bibinfo  {journal} {Advanced
  Photonics}\ }\textbf {\bibinfo {volume} {5}},\ \href
  {https://doi.org/10.1117/1.AP.5.6.066007} {10.1117/1.AP.5.6.066007} (\bibinfo
  {year} {2023})\BibitemShut {NoStop}%
\bibitem [{\citenamefont {Umucal{\ifmmode \imath \else \i \fi{}}lar}\ and\
  \citenamefont {Carusotto}(2012)}]{Umucalilar2012}%
  \BibitemOpen
  \bibfield  {author} {\bibinfo {author} {\bibfnamefont {R.~O.}\ \bibnamefont
  {Umucal{\ifmmode \imath \else \i \fi{}}lar}}\ and\ \bibinfo {author}
  {\bibfnamefont {I.}~\bibnamefont {Carusotto}},\ }\bibfield  {title} {\bibinfo
  {title} {Fractional quantum {H}all states of photons in an array of
  dissipative coupled cavities},\ }\href
  {https://doi.org/10.1103/PhysRevLett.108.206809} {\bibfield  {journal}
  {\bibinfo  {journal} {Phys. Rev. Lett.}\ }\textbf {\bibinfo {volume} {108}},\
  \bibinfo {pages} {206809} (\bibinfo {year} {2012})}\BibitemShut {NoStop}%
\bibitem [{\citenamefont {Hafezi}\ \emph {et~al.}(2011)\citenamefont {Hafezi},
  \citenamefont {Demler}, \citenamefont {Lukin},\ and\ \citenamefont
  {Taylor}}]{Hafezi2011}%
  \BibitemOpen
  \bibfield  {author} {\bibinfo {author} {\bibfnamefont {M.}~\bibnamefont
  {Hafezi}}, \bibinfo {author} {\bibfnamefont {E.~A.}\ \bibnamefont {Demler}},
  \bibinfo {author} {\bibfnamefont {M.~D.}\ \bibnamefont {Lukin}},\ and\
  \bibinfo {author} {\bibfnamefont {J.~M.}\ \bibnamefont {Taylor}},\ }\bibfield
   {title} {\bibinfo {title} {Robust optical delay lines with topological
  protection},\ }\href {https://doi.org/10.1038/nphys2063} {\bibfield
  {journal} {\bibinfo  {journal} {Nature Phys.}\ }\textbf {\bibinfo {volume}
  {7}},\ \bibinfo {pages} {907} (\bibinfo {year} {2011})}\BibitemShut {NoStop}%
\bibitem [{\citenamefont {Greentree}\ \emph {et~al.}(2006)\citenamefont
  {Greentree}, \citenamefont {Tahan}, \citenamefont {Cole},\ and\ \citenamefont
  {Hollenberg}}]{Greentree2006}%
  \BibitemOpen
  \bibfield  {author} {\bibinfo {author} {\bibfnamefont {A.~D.}\ \bibnamefont
  {Greentree}}, \bibinfo {author} {\bibfnamefont {C.}~\bibnamefont {Tahan}},
  \bibinfo {author} {\bibfnamefont {J.~H.}\ \bibnamefont {Cole}},\ and\
  \bibinfo {author} {\bibfnamefont {L.~C.~L.}\ \bibnamefont {Hollenberg}},\
  }\bibfield  {title} {\bibinfo {title} {Quantum phase transitions of light},\
  }\href {https://doi.org/10.1038/nphys466} {\bibfield  {journal} {\bibinfo
  {journal} {Nature Phys.}\ }\textbf {\bibinfo {volume} {2}},\ \bibinfo {pages}
  {856} (\bibinfo {year} {2006})}\BibitemShut {NoStop}%
\bibitem [{\citenamefont {Kavokin}\ \emph {et~al.}(2017)\citenamefont
  {Kavokin}, \citenamefont {Baumberg}, \citenamefont {Malpuech},\ and\
  \citenamefont {Laussy}}]{Kavokin2017_OxfPr}%
  \BibitemOpen
  \bibfield  {author} {\bibinfo {author} {\bibfnamefont {A.~V.}\ \bibnamefont
  {Kavokin}}, \bibinfo {author} {\bibfnamefont {J.~J.}\ \bibnamefont
  {Baumberg}}, \bibinfo {author} {\bibfnamefont {G.}~\bibnamefont {Malpuech}},\
  and\ \bibinfo {author} {\bibfnamefont {F.~P.}\ \bibnamefont {Laussy}},\
  }\href {https://doi.org/10.1093/oso/9780198782995.001.0001} {\emph {\bibinfo
  {title} {Microcavities}}}\ (\bibinfo  {publisher} {Oxford University Press},\
  \bibinfo {year} {2017})\BibitemShut {NoStop}%
\bibitem [{\citenamefont {Carusotto}\ and\ \citenamefont
  {Ciuti}(2013)}]{Carusotto2013}%
  \BibitemOpen
  \bibfield  {author} {\bibinfo {author} {\bibfnamefont {I.}~\bibnamefont
  {Carusotto}}\ and\ \bibinfo {author} {\bibfnamefont {C.}~\bibnamefont
  {Ciuti}},\ }\bibfield  {title} {\bibinfo {title} {Quantum fluids of light},\
  }\href {https://doi.org/10.1103/RevModPhys.85.299} {\bibfield  {journal}
  {\bibinfo  {journal} {Rev. Mod. Phys.}\ }\textbf {\bibinfo {volume} {85}},\
  \bibinfo {pages} {299} (\bibinfo {year} {2013})}\BibitemShut {NoStop}%
\bibitem [{\citenamefont {Ballarini}\ \emph {et~al.}(2017)\citenamefont
  {Ballarini}, \citenamefont {Caputo}, \citenamefont {Mu\~noz}, \citenamefont
  {De~Giorgi}, \citenamefont {Dominici}, \citenamefont
  {Szyma\ifmmode~\acute{n}\else \'{n}\fi{}ska}, \citenamefont {West},
  \citenamefont {Pfeiffer}, \citenamefont {Gigli}, \citenamefont {Laussy},\
  and\ \citenamefont {Sanvitto}}]{Ballarini2017}%
  \BibitemOpen
  \bibfield  {author} {\bibinfo {author} {\bibfnamefont {D.}~\bibnamefont
  {Ballarini}}, \bibinfo {author} {\bibfnamefont {D.}~\bibnamefont {Caputo}},
  \bibinfo {author} {\bibfnamefont {C.~S.}\ \bibnamefont {Mu\~noz}}, \bibinfo
  {author} {\bibfnamefont {M.}~\bibnamefont {De~Giorgi}}, \bibinfo {author}
  {\bibfnamefont {L.}~\bibnamefont {Dominici}}, \bibinfo {author}
  {\bibfnamefont {M.~H.}\ \bibnamefont {Szyma\ifmmode~\acute{n}\else
  \'{n}\fi{}ska}}, \bibinfo {author} {\bibfnamefont {K.}~\bibnamefont {West}},
  \bibinfo {author} {\bibfnamefont {L.~N.}\ \bibnamefont {Pfeiffer}}, \bibinfo
  {author} {\bibfnamefont {G.}~\bibnamefont {Gigli}}, \bibinfo {author}
  {\bibfnamefont {F.~P.}\ \bibnamefont {Laussy}},\ and\ \bibinfo {author}
  {\bibfnamefont {D.}~\bibnamefont {Sanvitto}},\ }\bibfield  {title} {\bibinfo
  {title} {Macroscopic two-dimensional polariton condensates},\ }\href
  {https://doi.org/10.1103/PhysRevLett.118.215301} {\bibfield  {journal}
  {\bibinfo  {journal} {Phys. Rev. Lett.}\ }\textbf {\bibinfo {volume} {118}},\
  \bibinfo {pages} {215301} (\bibinfo {year} {2017})}\BibitemShut {NoStop}%
\bibitem [{\citenamefont {Kasprzak}\ \emph {et~al.}(2006)\citenamefont
  {Kasprzak}, \citenamefont {Richard}, \citenamefont {Kundermann},
  \citenamefont {Baas}, \citenamefont {Jeambrun}, \citenamefont {Marchetti},
  \citenamefont {Szymanska}, \citenamefont {Andre}, \citenamefont {Staehli},
  \citenamefont {Savona}, \citenamefont {Littlewood}, \citenamefont {Deveaud},\
  and\ \citenamefont {Si~Dang}}]{Kasprzak2006}%
  \BibitemOpen
  \bibfield  {author} {\bibinfo {author} {\bibfnamefont {J.}~\bibnamefont
  {Kasprzak}}, \bibinfo {author} {\bibfnamefont {M.}~\bibnamefont {Richard}},
  \bibinfo {author} {\bibfnamefont {S.}~\bibnamefont {Kundermann}}, \bibinfo
  {author} {\bibfnamefont {A.}~\bibnamefont {Baas}}, \bibinfo {author}
  {\bibfnamefont {J.~M.~J.}\ \bibnamefont {Jeambrun}, \bibfnamefont
  {P.~Keeling}}, \bibinfo {author} {\bibfnamefont {F.~M.}\ \bibnamefont
  {Marchetti}}, \bibinfo {author} {\bibfnamefont {M.~H.}\ \bibnamefont
  {Szymanska}}, \bibinfo {author} {\bibfnamefont {R.}~\bibnamefont {Andre}},
  \bibinfo {author} {\bibfnamefont {J.~M.}\ \bibnamefont {Staehli}}, \bibinfo
  {author} {\bibfnamefont {V.}~\bibnamefont {Savona}}, \bibinfo {author}
  {\bibfnamefont {P.~B.}\ \bibnamefont {Littlewood}}, \bibinfo {author}
  {\bibfnamefont {B.}~\bibnamefont {Deveaud}},\ and\ \bibinfo {author}
  {\bibfnamefont {L.}~\bibnamefont {Si~Dang}},\ }\bibfield  {title} {\bibinfo
  {title} {Bose–{E}instein condensation of exciton polaritons},\ }\href
  {https://doi.org/10.1038/nature05131} {\bibfield  {journal} {\bibinfo
  {journal} {Nature}\ }\textbf {\bibinfo {volume} {443}},\ \bibinfo {pages}
  {409} (\bibinfo {year} {2006})}\BibitemShut {NoStop}%
\bibitem [{\citenamefont {Balili}\ \emph {et~al.}(2007)\citenamefont {Balili},
  \citenamefont {Hartwell}, \citenamefont {Snoke},\ and\ \citenamefont
  {West}}]{Balili2007}%
  \BibitemOpen
  \bibfield  {author} {\bibinfo {author} {\bibfnamefont {R.}~\bibnamefont
  {Balili}}, \bibinfo {author} {\bibfnamefont {V.}~\bibnamefont {Hartwell}},
  \bibinfo {author} {\bibfnamefont {D.}~\bibnamefont {Snoke}},\ and\ \bibinfo
  {author} {\bibfnamefont {K.}~\bibnamefont {West}},\ }\bibfield  {title}
  {\bibinfo {title} {Bose-{E}instein condensation of microcavity polaritons in
  a trap},\ }\href {https://doi.org/10.1126/science.1140990} {\bibfield
  {journal} {\bibinfo  {journal} {Science}\ }\textbf {\bibinfo {volume}
  {316}},\ \bibinfo {pages} {1007} (\bibinfo {year} {2007})}\BibitemShut
  {NoStop}%
\bibitem [{\citenamefont {Schneider}\ \emph {et~al.}(2013)\citenamefont
  {Schneider}, \citenamefont {Rahimi-Iman}, \citenamefont {Kim}, \citenamefont
  {Fischer}, \citenamefont {Savenko}, \citenamefont {Amthor}, \citenamefont
  {Lermer}, \citenamefont {Wolf}, \citenamefont {Worschech}, \citenamefont
  {Kulakovskii}, \citenamefont {Shelykh}, \citenamefont {Kamp}, \citenamefont
  {Reitzenstein}, \citenamefont {Forchel}, \citenamefont {Yamamoto},\ and\
  \citenamefont {Hofling}}]{Schneider2013}%
  \BibitemOpen
  \bibfield  {author} {\bibinfo {author} {\bibfnamefont {C.}~\bibnamefont
  {Schneider}}, \bibinfo {author} {\bibfnamefont {A.}~\bibnamefont
  {Rahimi-Iman}}, \bibinfo {author} {\bibfnamefont {N.~Y.}\ \bibnamefont
  {Kim}}, \bibinfo {author} {\bibfnamefont {J.}~\bibnamefont {Fischer}},
  \bibinfo {author} {\bibfnamefont {I.~G.}\ \bibnamefont {Savenko}}, \bibinfo
  {author} {\bibfnamefont {M.}~\bibnamefont {Amthor}}, \bibinfo {author}
  {\bibfnamefont {M.}~\bibnamefont {Lermer}}, \bibinfo {author} {\bibfnamefont
  {A.}~\bibnamefont {Wolf}}, \bibinfo {author} {\bibfnamefont {L.}~\bibnamefont
  {Worschech}}, \bibinfo {author} {\bibfnamefont {V.~D.}\ \bibnamefont
  {Kulakovskii}}, \bibinfo {author} {\bibfnamefont {I.~A.}\ \bibnamefont
  {Shelykh}}, \bibinfo {author} {\bibfnamefont {M.}~\bibnamefont {Kamp}},
  \bibinfo {author} {\bibfnamefont {S.}~\bibnamefont {Reitzenstein}}, \bibinfo
  {author} {\bibfnamefont {A.}~\bibnamefont {Forchel}}, \bibinfo {author}
  {\bibfnamefont {Y.}~\bibnamefont {Yamamoto}},\ and\ \bibinfo {author}
  {\bibfnamefont {S.}~\bibnamefont {Hofling}},\ }\bibfield  {title} {\bibinfo
  {title} {An electrically pumped polariton laser},\ }\href
  {https://doi.org/doi.org/10.1038/nature12036} {\bibfield  {journal} {\bibinfo
   {journal} {Nature}\ }\textbf {\bibinfo {volume} {497}},\ \bibinfo {pages}
  {348} (\bibinfo {year} {2013})}\BibitemShut {NoStop}%
\bibitem [{\citenamefont {Baboux}\ \emph {et~al.}(2016)\citenamefont {Baboux},
  \citenamefont {Ge}, \citenamefont {Jacqmin}, \citenamefont {Biondi},
  \citenamefont {Galopin}, \citenamefont {Lema\^{\i}tre}, \citenamefont
  {Le~Gratiet}, \citenamefont {Sagnes}, \citenamefont {Schmidt}, \citenamefont
  {T\"ureci}, \citenamefont {Amo},\ and\ \citenamefont
  {Bloch}}]{PhysRevLett.116.066402}%
  \BibitemOpen
  \bibfield  {author} {\bibinfo {author} {\bibfnamefont {F.}~\bibnamefont
  {Baboux}}, \bibinfo {author} {\bibfnamefont {L.}~\bibnamefont {Ge}}, \bibinfo
  {author} {\bibfnamefont {T.}~\bibnamefont {Jacqmin}}, \bibinfo {author}
  {\bibfnamefont {M.}~\bibnamefont {Biondi}}, \bibinfo {author} {\bibfnamefont
  {E.}~\bibnamefont {Galopin}}, \bibinfo {author} {\bibfnamefont
  {A.}~\bibnamefont {Lema\^{\i}tre}}, \bibinfo {author} {\bibfnamefont
  {L.}~\bibnamefont {Le~Gratiet}}, \bibinfo {author} {\bibfnamefont
  {I.}~\bibnamefont {Sagnes}}, \bibinfo {author} {\bibfnamefont
  {S.}~\bibnamefont {Schmidt}}, \bibinfo {author} {\bibfnamefont {H.~E.}\
  \bibnamefont {T\"ureci}}, \bibinfo {author} {\bibfnamefont {A.}~\bibnamefont
  {Amo}},\ and\ \bibinfo {author} {\bibfnamefont {J.}~\bibnamefont {Bloch}},\
  }\bibfield  {title} {\bibinfo {title} {Bosonic condensation and
  disorder-induced localization in a flat band},\ }\href
  {https://doi.org/10.1103/PhysRevLett.116.066402} {\bibfield  {journal}
  {\bibinfo  {journal} {Phys. Rev. Lett.}\ }\textbf {\bibinfo {volume} {116}},\
  \bibinfo {pages} {066402} (\bibinfo {year} {2016})}\BibitemShut {NoStop}%
\bibitem [{\citenamefont {Mili\ifmmode \acute{c}\else
  \'{c}\fi{}evi\ifmmode~\acute{c}\else \'{c}\fi{}}\ \emph
  {et~al.}(2017)\citenamefont {Mili\ifmmode \acute{c}\else
  \'{c}\fi{}evi\ifmmode~\acute{c}\else \'{c}\fi{}}, \citenamefont {Ozawa},
  \citenamefont {Montambaux}, \citenamefont {Carusotto}, \citenamefont
  {Galopin}, \citenamefont {Lema\^{\i}tre}, \citenamefont {Le~Gratiet},
  \citenamefont {Sagnes}, \citenamefont {Bloch},\ and\ \citenamefont
  {Amo}}]{Milicevic2017}%
  \BibitemOpen
  \bibfield  {author} {\bibinfo {author} {\bibfnamefont {M.}~\bibnamefont
  {Mili\ifmmode \acute{c}\else \'{c}\fi{}evi\ifmmode~\acute{c}\else
  \'{c}\fi{}}}, \bibinfo {author} {\bibfnamefont {T.}~\bibnamefont {Ozawa}},
  \bibinfo {author} {\bibfnamefont {G.}~\bibnamefont {Montambaux}}, \bibinfo
  {author} {\bibfnamefont {I.}~\bibnamefont {Carusotto}}, \bibinfo {author}
  {\bibfnamefont {E.}~\bibnamefont {Galopin}}, \bibinfo {author} {\bibfnamefont
  {A.}~\bibnamefont {Lema\^{\i}tre}}, \bibinfo {author} {\bibfnamefont
  {L.}~\bibnamefont {Le~Gratiet}}, \bibinfo {author} {\bibfnamefont
  {I.}~\bibnamefont {Sagnes}}, \bibinfo {author} {\bibfnamefont
  {J.}~\bibnamefont {Bloch}},\ and\ \bibinfo {author} {\bibfnamefont
  {A.}~\bibnamefont {Amo}},\ }\bibfield  {title} {\bibinfo {title} {Orbital
  edge states in a photonic honeycomb lattice},\ }\href
  {https://doi.org/10.1103/PhysRevLett.118.107403} {\bibfield  {journal}
  {\bibinfo  {journal} {Phys. Rev. Lett.}\ }\textbf {\bibinfo {volume} {118}},\
  \bibinfo {pages} {107403} (\bibinfo {year} {2017})}\BibitemShut {NoStop}%
\bibitem [{\citenamefont {Gao}\ \emph {et~al.}(2018)\citenamefont {Gao},
  \citenamefont {Egorov}, \citenamefont {Estrecho}, \citenamefont {Winkler},
  \citenamefont {Kamp}, \citenamefont {Schneider}, \citenamefont {H\"ofling},
  \citenamefont {Truscott},\ and\ \citenamefont {Ostrovskaya}}]{Gao2018}%
  \BibitemOpen
  \bibfield  {author} {\bibinfo {author} {\bibfnamefont {T.}~\bibnamefont
  {Gao}}, \bibinfo {author} {\bibfnamefont {O.~A.}\ \bibnamefont {Egorov}},
  \bibinfo {author} {\bibfnamefont {E.}~\bibnamefont {Estrecho}}, \bibinfo
  {author} {\bibfnamefont {K.}~\bibnamefont {Winkler}}, \bibinfo {author}
  {\bibfnamefont {M.}~\bibnamefont {Kamp}}, \bibinfo {author} {\bibfnamefont
  {C.}~\bibnamefont {Schneider}}, \bibinfo {author} {\bibfnamefont
  {S.}~\bibnamefont {H\"ofling}}, \bibinfo {author} {\bibfnamefont {A.~G.}\
  \bibnamefont {Truscott}},\ and\ \bibinfo {author} {\bibfnamefont {E.~A.}\
  \bibnamefont {Ostrovskaya}},\ }\bibfield  {title} {\bibinfo {title}
  {Controlled ordering of topological charges in an exciton-polariton chain},\
  }\href {https://doi.org/10.1103/PhysRevLett.121.225302} {\bibfield  {journal}
  {\bibinfo  {journal} {Phys. Rev. Lett.}\ }\textbf {\bibinfo {volume} {121}},\
  \bibinfo {pages} {225302} (\bibinfo {year} {2018})}\BibitemShut {NoStop}%
\bibitem [{\citenamefont {Whittaker}\ \emph {et~al.}(2018)\citenamefont
  {Whittaker}, \citenamefont {Cancellieri}, \citenamefont {Walker},
  \citenamefont {Gulevich}, \citenamefont {Schomerus}, \citenamefont
  {Vaitiekus}, \citenamefont {Royall}, \citenamefont {Whittaker}, \citenamefont
  {Clarke}, \citenamefont {Iorsh}, \citenamefont {Shelykh}, \citenamefont
  {Skolnick},\ and\ \citenamefont {Krizhanovskii}}]{Whittaker2018}%
  \BibitemOpen
  \bibfield  {author} {\bibinfo {author} {\bibfnamefont {C.~E.}\ \bibnamefont
  {Whittaker}}, \bibinfo {author} {\bibfnamefont {E.}~\bibnamefont
  {Cancellieri}}, \bibinfo {author} {\bibfnamefont {P.~M.}\ \bibnamefont
  {Walker}}, \bibinfo {author} {\bibfnamefont {D.~R.}\ \bibnamefont
  {Gulevich}}, \bibinfo {author} {\bibfnamefont {H.}~\bibnamefont {Schomerus}},
  \bibinfo {author} {\bibfnamefont {D.}~\bibnamefont {Vaitiekus}}, \bibinfo
  {author} {\bibfnamefont {B.}~\bibnamefont {Royall}}, \bibinfo {author}
  {\bibfnamefont {D.~M.}\ \bibnamefont {Whittaker}}, \bibinfo {author}
  {\bibfnamefont {E.}~\bibnamefont {Clarke}}, \bibinfo {author} {\bibfnamefont
  {I.~V.}\ \bibnamefont {Iorsh}}, \bibinfo {author} {\bibfnamefont {I.~A.}\
  \bibnamefont {Shelykh}}, \bibinfo {author} {\bibfnamefont {M.~S.}\
  \bibnamefont {Skolnick}},\ and\ \bibinfo {author} {\bibfnamefont {D.~N.}\
  \bibnamefont {Krizhanovskii}},\ }\bibfield  {title} {\bibinfo {title}
  {Exciton polaritons in a two-dimensional {L}ieb lattice with spin-orbit
  coupling},\ }\href {https://doi.org/10.1103/PhysRevLett.120.097401}
  {\bibfield  {journal} {\bibinfo  {journal} {Phys. Rev. Lett.}\ }\textbf
  {\bibinfo {volume} {120}},\ \bibinfo {pages} {097401} (\bibinfo {year}
  {2018})}\BibitemShut {NoStop}%
\bibitem [{\citenamefont {Whittaker}\ \emph {et~al.}(2021)\citenamefont
  {Whittaker}, \citenamefont {Dowling}, \citenamefont {Nalitov}, \citenamefont
  {Yulin}, \citenamefont {Royall}, \citenamefont {Clarke}, \citenamefont
  {Skolnick}, \citenamefont {Shelykh},\ and\ \citenamefont
  {Krizhanovskii}}]{Whittaker2021}%
  \BibitemOpen
  \bibfield  {author} {\bibinfo {author} {\bibfnamefont {C.~E.}\ \bibnamefont
  {Whittaker}}, \bibinfo {author} {\bibfnamefont {T.}~\bibnamefont {Dowling}},
  \bibinfo {author} {\bibfnamefont {A.~V.}\ \bibnamefont {Nalitov}}, \bibinfo
  {author} {\bibfnamefont {A.~V.}\ \bibnamefont {Yulin}}, \bibinfo {author}
  {\bibfnamefont {B.}~\bibnamefont {Royall}}, \bibinfo {author} {\bibfnamefont
  {E.}~\bibnamefont {Clarke}}, \bibinfo {author} {\bibfnamefont {M.~S.}\
  \bibnamefont {Skolnick}}, \bibinfo {author} {\bibfnamefont {I.~A.}\
  \bibnamefont {Shelykh}},\ and\ \bibinfo {author} {\bibfnamefont {D.~N.}\
  \bibnamefont {Krizhanovskii}},\ }\bibfield  {title} {\bibinfo {title}
  {Optical analogue of {D}resselhaus spin–orbit interaction in photonic
  graphene},\ }\href {https://doi.org/10.1038/s41566-020-00729-z} {\bibfield
  {journal} {\bibinfo  {journal} {Nat. Photon.}\ }\textbf {\bibinfo {volume}
  {15}},\ \bibinfo {pages} {193} (\bibinfo {year} {2021})}\BibitemShut
  {NoStop}%
\bibitem [{\citenamefont {Kozin}\ \emph {et~al.}(2018)\citenamefont {Kozin},
  \citenamefont {Shelykh}, \citenamefont {Nalitov},\ and\ \citenamefont
  {Iorsh}}]{Kozin2018}%
  \BibitemOpen
  \bibfield  {author} {\bibinfo {author} {\bibfnamefont {V.~K.}\ \bibnamefont
  {Kozin}}, \bibinfo {author} {\bibfnamefont {I.~A.}\ \bibnamefont {Shelykh}},
  \bibinfo {author} {\bibfnamefont {A.~V.}\ \bibnamefont {Nalitov}},\ and\
  \bibinfo {author} {\bibfnamefont {I.~V.}\ \bibnamefont {Iorsh}},\ }\bibfield
  {title} {\bibinfo {title} {Topological metamaterials based on polariton
  rings},\ }\href {https://doi.org/10.1103/PhysRevB.98.125115} {\bibfield
  {journal} {\bibinfo  {journal} {Phys. Rev. B}\ }\textbf {\bibinfo {volume}
  {98}},\ \bibinfo {pages} {125115} (\bibinfo {year} {2018})}\BibitemShut
  {NoStop}%
\bibitem [{\citenamefont {Ghosh}\ and\ \citenamefont
  {Liew}(2020)}]{Ghosh2020_QI}%
  \BibitemOpen
  \bibfield  {author} {\bibinfo {author} {\bibfnamefont {S.}~\bibnamefont
  {Ghosh}}\ and\ \bibinfo {author} {\bibfnamefont {T.~C.~H.}\ \bibnamefont
  {Liew}},\ }\bibfield  {title} {\bibinfo {title} {Quantum computing with
  exciton-polariton condensates},\ }\href
  {https://doi.org/10.1038/s41534-020-0244-x} {\bibfield  {journal} {\bibinfo
  {journal} {npj Quantum Inf.}\ }\textbf {\bibinfo {volume} {6}},\ \bibinfo
  {pages} {16} (\bibinfo {year} {2020})}\BibitemShut {NoStop}%
\bibitem [{\citenamefont {Kavokin}\ \emph {et~al.}(2022)\citenamefont
  {Kavokin}, \citenamefont {Liew}, \citenamefont {Schneider}, \citenamefont
  {Lagoudakis}, \citenamefont {Klembt},\ and\ \citenamefont
  {Hoefling}}]{Kavokin2022}%
  \BibitemOpen
  \bibfield  {author} {\bibinfo {author} {\bibfnamefont {A.}~\bibnamefont
  {Kavokin}}, \bibinfo {author} {\bibfnamefont {T.~C.~H.}\ \bibnamefont
  {Liew}}, \bibinfo {author} {\bibfnamefont {C.}~\bibnamefont {Schneider}},
  \bibinfo {author} {\bibfnamefont {P.~G.}\ \bibnamefont {Lagoudakis}},
  \bibinfo {author} {\bibfnamefont {S.}~\bibnamefont {Klembt}},\ and\ \bibinfo
  {author} {\bibfnamefont {S.}~\bibnamefont {Hoefling}},\ }\bibfield  {title}
  {\bibinfo {title} {Polariton condensates for classical and quantum
  computing},\ }\href {https://doi.org/10.1038/s42254-022-00447-1} {\bibfield
  {journal} {\bibinfo  {journal} {Nat. Rev. Phys.}\ }\textbf {\bibinfo {volume}
  {4}},\ \bibinfo {pages} {435} (\bibinfo {year} {2022})}\BibitemShut {NoStop}%
\bibitem [{\citenamefont {Kalinin}\ and\ \citenamefont
  {Berloff}(2018)}]{Kalinin2018}%
  \BibitemOpen
  \bibfield  {author} {\bibinfo {author} {\bibfnamefont {K.~P.}\ \bibnamefont
  {Kalinin}}\ and\ \bibinfo {author} {\bibfnamefont {N.~G.}\ \bibnamefont
  {Berloff}},\ }\bibfield  {title} {\bibinfo {title} {Simulating {I}sing and
  $n$-state planar {P}otts models and external fields with nonequilibrium
  condensates},\ }\href {https://doi.org/10.1103/PhysRevLett.121.235302}
  {\bibfield  {journal} {\bibinfo  {journal} {Phys. Rev. Lett.}\ }\textbf
  {\bibinfo {volume} {121}},\ \bibinfo {pages} {235302} (\bibinfo {year}
  {2018})}\BibitemShut {NoStop}%
\bibitem [{\citenamefont {Suchomel}\ \emph {et~al.}(2018)\citenamefont
  {Suchomel}, \citenamefont {Klembt}, \citenamefont {Harder}, \citenamefont
  {Klaas}, \citenamefont {Egorov}, \citenamefont {Winkler}, \citenamefont
  {Emmerling}, \citenamefont {Thomale}, \citenamefont {H\"ofling},\ and\
  \citenamefont {Schneider}}]{Suchomel2018}%
  \BibitemOpen
  \bibfield  {author} {\bibinfo {author} {\bibfnamefont {H.}~\bibnamefont
  {Suchomel}}, \bibinfo {author} {\bibfnamefont {S.}~\bibnamefont {Klembt}},
  \bibinfo {author} {\bibfnamefont {T.~H.}\ \bibnamefont {Harder}}, \bibinfo
  {author} {\bibfnamefont {M.}~\bibnamefont {Klaas}}, \bibinfo {author}
  {\bibfnamefont {O.~A.}\ \bibnamefont {Egorov}}, \bibinfo {author}
  {\bibfnamefont {K.}~\bibnamefont {Winkler}}, \bibinfo {author} {\bibfnamefont
  {M.}~\bibnamefont {Emmerling}}, \bibinfo {author} {\bibfnamefont
  {R.}~\bibnamefont {Thomale}}, \bibinfo {author} {\bibfnamefont
  {S.}~\bibnamefont {H\"ofling}},\ and\ \bibinfo {author} {\bibfnamefont
  {C.}~\bibnamefont {Schneider}},\ }\bibfield  {title} {\bibinfo {title}
  {Platform for electrically pumped polariton simulators and topological
  lasers},\ }\href {https://doi.org/10.1103/PhysRevLett.121.257402} {\bibfield
  {journal} {\bibinfo  {journal} {Phys. Rev. Lett.}\ }\textbf {\bibinfo
  {volume} {121}},\ \bibinfo {pages} {257402} (\bibinfo {year}
  {2018})}\BibitemShut {NoStop}%
\bibitem [{\citenamefont {Opala}\ \emph {et~al.}(2019)\citenamefont {Opala},
  \citenamefont {Ghosh}, \citenamefont {Liew},\ and\ \citenamefont
  {Matuszewski}}]{Opala2019}%
  \BibitemOpen
  \bibfield  {author} {\bibinfo {author} {\bibfnamefont {A.}~\bibnamefont
  {Opala}}, \bibinfo {author} {\bibfnamefont {S.}~\bibnamefont {Ghosh}},
  \bibinfo {author} {\bibfnamefont {T.~C.}\ \bibnamefont {Liew}},\ and\
  \bibinfo {author} {\bibfnamefont {M.}~\bibnamefont {Matuszewski}},\
  }\bibfield  {title} {\bibinfo {title} {Neuromorphic computing in
  {G}inzburg-{L}andau polariton-lattice systems},\ }\href
  {https://doi.org/10.1103/PhysRevApplied.11.064029} {\bibfield  {journal}
  {\bibinfo  {journal} {Phys. Rev. Applied}\ }\textbf {\bibinfo {volume}
  {11}},\ \bibinfo {pages} {064029} (\bibinfo {year} {2019})}\BibitemShut
  {NoStop}%
\bibitem [{\citenamefont {Boulier}\ \emph {et~al.}(2020)\citenamefont
  {Boulier}, \citenamefont {Jacquet}, \citenamefont {Maître}, \citenamefont
  {Lerario}, \citenamefont {Claude}, \citenamefont {Pigeon}, \citenamefont
  {Glorieux}, \citenamefont {Amo}, \citenamefont {Bloch}, \citenamefont
  {Bramati},\ and\ \citenamefont {Giacobino}}]{Boulier2020_AQT}%
  \BibitemOpen
  \bibfield  {author} {\bibinfo {author} {\bibfnamefont {T.}~\bibnamefont
  {Boulier}}, \bibinfo {author} {\bibfnamefont {M.~J.}\ \bibnamefont
  {Jacquet}}, \bibinfo {author} {\bibfnamefont {A.}~\bibnamefont {Maître}},
  \bibinfo {author} {\bibfnamefont {G.}~\bibnamefont {Lerario}}, \bibinfo
  {author} {\bibfnamefont {F.}~\bibnamefont {Claude}}, \bibinfo {author}
  {\bibfnamefont {S.}~\bibnamefont {Pigeon}}, \bibinfo {author} {\bibfnamefont
  {Q.}~\bibnamefont {Glorieux}}, \bibinfo {author} {\bibfnamefont
  {A.}~\bibnamefont {Amo}}, \bibinfo {author} {\bibfnamefont {J.}~\bibnamefont
  {Bloch}}, \bibinfo {author} {\bibfnamefont {A.}~\bibnamefont {Bramati}},\
  and\ \bibinfo {author} {\bibfnamefont {E.}~\bibnamefont {Giacobino}},\
  }\bibfield  {title} {\bibinfo {title} {Microcavity polaritons for quantum
  simulation},\ }\href {https://doi.org/https://doi.org/10.1002/qute.202000052}
  {\bibfield  {journal} {\bibinfo  {journal} {Adv. Quantum Technol.}\ }\textbf
  {\bibinfo {volume} {3}},\ \bibinfo {pages} {2000052} (\bibinfo {year}
  {2020})}\BibitemShut {NoStop}%
\bibitem [{\citenamefont {Marcucci}\ \emph {et~al.}(2020)\citenamefont
  {Marcucci}, \citenamefont {Pierangeli},\ and\ \citenamefont
  {Conti}}]{Marcucci2020}%
  \BibitemOpen
  \bibfield  {author} {\bibinfo {author} {\bibfnamefont {G.}~\bibnamefont
  {Marcucci}}, \bibinfo {author} {\bibfnamefont {D.}~\bibnamefont
  {Pierangeli}},\ and\ \bibinfo {author} {\bibfnamefont {C.}~\bibnamefont
  {Conti}},\ }\bibfield  {title} {\bibinfo {title} {Theory of neuromorphic
  computing by waves: Machine learning by rogue waves, dispersive shocks, and
  solitons},\ }\href {https://doi.org/10.1103/PhysRevLett.125.093901}
  {\bibfield  {journal} {\bibinfo  {journal} {Phys. Rev. Lett.}\ }\textbf
  {\bibinfo {volume} {125}},\ \bibinfo {pages} {093901} (\bibinfo {year}
  {2020})}\BibitemShut {NoStop}%
\bibitem [{\citenamefont {Shelykh}\ \emph {et~al.}(2010)\citenamefont
  {Shelykh}, \citenamefont {Kavokin}, \citenamefont {Rubo}, \citenamefont
  {Liew},\ and\ \citenamefont {Malpuech}}]{ShelykhReview}%
  \BibitemOpen
  \bibfield  {author} {\bibinfo {author} {\bibfnamefont {I.~A.}\ \bibnamefont
  {Shelykh}}, \bibinfo {author} {\bibfnamefont {A.~V.}\ \bibnamefont
  {Kavokin}}, \bibinfo {author} {\bibfnamefont {Y.~G.}\ \bibnamefont {Rubo}},
  \bibinfo {author} {\bibfnamefont {T.~C.~H.}\ \bibnamefont {Liew}},\ and\
  \bibinfo {author} {\bibfnamefont {G.}~\bibnamefont {Malpuech}},\ }\bibfield
  {title} {\bibinfo {title} {Polariton polarization-sensitive phenomena in
  planar semiconductor microcavities},\ }\href
  {https://doi.org/10.1088/0268-1242/25/1/013001} {\bibfield  {journal}
  {\bibinfo  {journal} {Semicond. Sci. Technol.}\ }\textbf {\bibinfo {volume}
  {25}},\ \bibinfo {pages} {013001} (\bibinfo {year} {2010})}\BibitemShut
  {NoStop}%
\bibitem [{\citenamefont {Sala}\ \emph {et~al.}(2015)\citenamefont {Sala},
  \citenamefont {Solnyshkov}, \citenamefont {Carusotto}, \citenamefont
  {Jacqmin}, \citenamefont {Lemaître}, \citenamefont {Terças}, \citenamefont
  {Nalitov}, \citenamefont {Abbarchi}, \citenamefont {Galopin}, \citenamefont
  {Sagnes}, \citenamefont {Bloch}, \citenamefont {Malpuech},\ and\
  \citenamefont {Amo}}]{Sala2015}%
  \BibitemOpen
  \bibfield  {author} {\bibinfo {author} {\bibfnamefont {V.~G.}\ \bibnamefont
  {Sala}}, \bibinfo {author} {\bibfnamefont {D.~D.}\ \bibnamefont
  {Solnyshkov}}, \bibinfo {author} {\bibfnamefont {I.}~\bibnamefont
  {Carusotto}}, \bibinfo {author} {\bibfnamefont {T.}~\bibnamefont {Jacqmin}},
  \bibinfo {author} {\bibfnamefont {A.}~\bibnamefont {Lemaître}}, \bibinfo
  {author} {\bibfnamefont {H.}~\bibnamefont {Terças}}, \bibinfo {author}
  {\bibfnamefont {A.}~\bibnamefont {Nalitov}}, \bibinfo {author} {\bibfnamefont
  {M.}~\bibnamefont {Abbarchi}}, \bibinfo {author} {\bibfnamefont
  {E.}~\bibnamefont {Galopin}}, \bibinfo {author} {\bibfnamefont
  {I.}~\bibnamefont {Sagnes}}, \bibinfo {author} {\bibfnamefont
  {J.}~\bibnamefont {Bloch}}, \bibinfo {author} {\bibfnamefont
  {G.}~\bibnamefont {Malpuech}},\ and\ \bibinfo {author} {\bibfnamefont
  {A.}~\bibnamefont {Amo}},\ }\bibfield  {title} {\bibinfo {title} {Spin-orbit
  coupling for photons and polaritons in microstructures},\ }\href
  {https://doi.org/10.1103/PhysRevX.5.011034} {\bibfield  {journal} {\bibinfo
  {journal} {Phys. Rev. X}\ }\textbf {\bibinfo {volume} {5}},\ \bibinfo {pages}
  {011034} (\bibinfo {year} {2015})}\BibitemShut {NoStop}%
\bibitem [{\citenamefont {Nalitov}\ \emph {et~al.}(2015)\citenamefont
  {Nalitov}, \citenamefont {Solnyshkov},\ and\ \citenamefont
  {Malpuech}}]{Nalitov2015}%
  \BibitemOpen
  \bibfield  {author} {\bibinfo {author} {\bibfnamefont {A.~V.}\ \bibnamefont
  {Nalitov}}, \bibinfo {author} {\bibfnamefont {D.~D.}\ \bibnamefont
  {Solnyshkov}},\ and\ \bibinfo {author} {\bibfnamefont {G.}~\bibnamefont
  {Malpuech}},\ }\bibfield  {title} {\bibinfo {title} {Polariton $\mathbb{Z}$
  topological insulator},\ }\href
  {https://doi.org/10.1103/PhysRevLett.114.116401} {\bibfield  {journal}
  {\bibinfo  {journal} {Phys. Rev. Lett.}\ }\textbf {\bibinfo {volume} {114}},\
  \bibinfo {pages} {116401} (\bibinfo {year} {2015})}\BibitemShut {NoStop}%
\bibitem [{\citenamefont {Klembt}\ \emph {et~al.}(2018)\citenamefont {Klembt},
  \citenamefont {Harder}, \citenamefont {Egorov}, \citenamefont {Winkler},
  \citenamefont {Ge}, \citenamefont {Bandres}, \citenamefont {Emmerling},
  \citenamefont {Worschech}, \citenamefont {Liew}, \citenamefont {Segev},
  \citenamefont {Schneider},\ and\ \citenamefont
  {H{\"o}fling}}]{klembt2018exciton}%
  \BibitemOpen
  \bibfield  {author} {\bibinfo {author} {\bibfnamefont {S.}~\bibnamefont
  {Klembt}}, \bibinfo {author} {\bibfnamefont {T.~H.}\ \bibnamefont {Harder}},
  \bibinfo {author} {\bibfnamefont {O.~A.}\ \bibnamefont {Egorov}}, \bibinfo
  {author} {\bibfnamefont {K.}~\bibnamefont {Winkler}}, \bibinfo {author}
  {\bibfnamefont {R.}~\bibnamefont {Ge}}, \bibinfo {author} {\bibfnamefont
  {M.~A.}\ \bibnamefont {Bandres}}, \bibinfo {author} {\bibfnamefont
  {M.}~\bibnamefont {Emmerling}}, \bibinfo {author} {\bibfnamefont
  {L.}~\bibnamefont {Worschech}}, \bibinfo {author} {\bibfnamefont {T.~C.~H.}\
  \bibnamefont {Liew}}, \bibinfo {author} {\bibfnamefont {M.}~\bibnamefont
  {Segev}}, \bibinfo {author} {\bibfnamefont {C.}~\bibnamefont {Schneider}},\
  and\ \bibinfo {author} {\bibfnamefont {S.}~\bibnamefont {H{\"o}fling}},\
  }\bibfield  {title} {\bibinfo {title} {Exciton-polariton topological
  insulator},\ }\href {https://doi.org/10.1038/s41586-018-0601-5} {\bibfield
  {journal} {\bibinfo  {journal} {Nature}\ }\textbf {\bibinfo {volume} {562}},\
  \bibinfo {pages} {552} (\bibinfo {year} {2018})}\BibitemShut {NoStop}%
\bibitem [{\citenamefont {Gulevich}\ \emph {et~al.}(2016)\citenamefont
  {Gulevich}, \citenamefont {Yudin}, \citenamefont {Iorsh},\ and\ \citenamefont
  {Shelykh}}]{Gulevich-kagome}%
  \BibitemOpen
  \bibfield  {author} {\bibinfo {author} {\bibfnamefont {D.~R.}\ \bibnamefont
  {Gulevich}}, \bibinfo {author} {\bibfnamefont {D.}~\bibnamefont {Yudin}},
  \bibinfo {author} {\bibfnamefont {I.~V.}\ \bibnamefont {Iorsh}},\ and\
  \bibinfo {author} {\bibfnamefont {I.~A.}\ \bibnamefont {Shelykh}},\
  }\bibfield  {title} {\bibinfo {title} {Kagome lattice from an
  exciton-polariton perspective},\ }\href
  {https://doi.org/10.1103/PhysRevB.94.115437} {\bibfield  {journal} {\bibinfo
  {journal} {Phys. Rev. B}\ }\textbf {\bibinfo {volume} {94}},\ \bibinfo
  {pages} {115437} (\bibinfo {year} {2016})}\BibitemShut {NoStop}%
\bibitem [{\citenamefont {Vladimirova}\ \emph {et~al.}(2010)\citenamefont
  {Vladimirova}, \citenamefont {Cronenberger}, \citenamefont {Scalbert},
  \citenamefont {Kavokin}, \citenamefont {Miard}, \citenamefont {Lemaître},
  \citenamefont {Bloch}, \citenamefont {Solnyshkov}, \citenamefont {Malpuech},\
  and\ \citenamefont {Kavokin}}]{Vladimirova2010}%
  \BibitemOpen
  \bibfield  {author} {\bibinfo {author} {\bibfnamefont {M.}~\bibnamefont
  {Vladimirova}}, \bibinfo {author} {\bibfnamefont {S.}~\bibnamefont
  {Cronenberger}}, \bibinfo {author} {\bibfnamefont {D.}~\bibnamefont
  {Scalbert}}, \bibinfo {author} {\bibfnamefont {K.~V.}\ \bibnamefont
  {Kavokin}}, \bibinfo {author} {\bibfnamefont {A.}~\bibnamefont {Miard}},
  \bibinfo {author} {\bibfnamefont {A.}~\bibnamefont {Lemaître}}, \bibinfo
  {author} {\bibfnamefont {J.}~\bibnamefont {Bloch}}, \bibinfo {author}
  {\bibfnamefont {D.}~\bibnamefont {Solnyshkov}}, \bibinfo {author}
  {\bibfnamefont {G.}~\bibnamefont {Malpuech}},\ and\ \bibinfo {author}
  {\bibfnamefont {A.~V.}\ \bibnamefont {Kavokin}},\ }\bibfield  {title}
  {\bibinfo {title} {Polariton-polariton interaction constants in
  microcavities},\ }\href {https://doi.org/10.1103/PhysRevB.82.075301}
  {\bibfield  {journal} {\bibinfo  {journal} {Phys. Rev. B}\ }\textbf {\bibinfo
  {volume} {82}},\ \bibinfo {pages} {075301} (\bibinfo {year}
  {2010})}\BibitemShut {NoStop}%
\bibitem [{\citenamefont {Ciuti}\ \emph {et~al.}(1998)\citenamefont {Ciuti},
  \citenamefont {Savona}, \citenamefont {Piermarocchi}, \citenamefont
  {Quattropani},\ and\ \citenamefont {Schwendimann}}]{Ciuti1998}%
  \BibitemOpen
  \bibfield  {author} {\bibinfo {author} {\bibfnamefont {C.}~\bibnamefont
  {Ciuti}}, \bibinfo {author} {\bibfnamefont {V.}~\bibnamefont {Savona}},
  \bibinfo {author} {\bibfnamefont {C.}~\bibnamefont {Piermarocchi}}, \bibinfo
  {author} {\bibfnamefont {A.}~\bibnamefont {Quattropani}},\ and\ \bibinfo
  {author} {\bibfnamefont {P.}~\bibnamefont {Schwendimann}},\ }\bibfield
  {title} {\bibinfo {title} {Role of the exchange of carriers in elastic
  exciton-exciton scattering in quantum wells},\ }\href
  {https://doi.org/10.1103/PhysRevB.58.7926} {\bibfield  {journal} {\bibinfo
  {journal} {Phys. Rev. B}\ }\textbf {\bibinfo {volume} {58}},\ \bibinfo
  {pages} {7926} (\bibinfo {year} {1998})}\BibitemShut {NoStop}%
\bibitem [{\citenamefont {Glazov}\ \emph {et~al.}(2009)\citenamefont {Glazov},
  \citenamefont {Ouerdane}, \citenamefont {Pilozzi}, \citenamefont {Malpuech},
  \citenamefont {Kavokin},\ and\ \citenamefont {D'Andrea}}]{Glazov2009}%
  \BibitemOpen
  \bibfield  {author} {\bibinfo {author} {\bibfnamefont {M.~M.}\ \bibnamefont
  {Glazov}}, \bibinfo {author} {\bibfnamefont {H.}~\bibnamefont {Ouerdane}},
  \bibinfo {author} {\bibfnamefont {L.}~\bibnamefont {Pilozzi}}, \bibinfo
  {author} {\bibfnamefont {G.}~\bibnamefont {Malpuech}}, \bibinfo {author}
  {\bibfnamefont {A.~V.}\ \bibnamefont {Kavokin}},\ and\ \bibinfo {author}
  {\bibfnamefont {A.}~\bibnamefont {D'Andrea}},\ }\bibfield  {title} {\bibinfo
  {title} {Polariton-polariton scattering in microcavities: A microscopic
  theory},\ }\href {https://doi.org/10.1103/PhysRevB.80.155306} {\bibfield
  {journal} {\bibinfo  {journal} {Phys. Rev. B}\ }\textbf {\bibinfo {volume}
  {80}},\ \bibinfo {pages} {155306} (\bibinfo {year} {2009})}\BibitemShut
  {NoStop}%
\bibitem [{\citenamefont {Shelykh}\ \emph {et~al.}(2004)\citenamefont
  {Shelykh}, \citenamefont {Malpuech}, \citenamefont {Kavokin}, \citenamefont
  {Kavokin},\ and\ \citenamefont {Bigenwald}}]{Shelykh2004}%
  \BibitemOpen
  \bibfield  {author} {\bibinfo {author} {\bibfnamefont {I.}~\bibnamefont
  {Shelykh}}, \bibinfo {author} {\bibfnamefont {G.}~\bibnamefont {Malpuech}},
  \bibinfo {author} {\bibfnamefont {K.~V.}\ \bibnamefont {Kavokin}}, \bibinfo
  {author} {\bibfnamefont {A.~V.}\ \bibnamefont {Kavokin}},\ and\ \bibinfo
  {author} {\bibfnamefont {P.}~\bibnamefont {Bigenwald}},\ }\bibfield  {title}
  {\bibinfo {title} {Spin dynamics of interacting exciton polaritons in
  microcavities},\ }\href {https://doi.org/10.1103/PhysRevB.70.115301}
  {\bibfield  {journal} {\bibinfo  {journal} {Phys. Rev. B}\ }\textbf {\bibinfo
  {volume} {70}},\ \bibinfo {pages} {115301} (\bibinfo {year}
  {2004})}\BibitemShut {NoStop}%
\bibitem [{\citenamefont {Laussy}\ \emph {et~al.}(2006)\citenamefont {Laussy},
  \citenamefont {Shelykh}, \citenamefont {Malpuech},\ and\ \citenamefont
  {Kavokin}}]{Laussy2006}%
  \BibitemOpen
  \bibfield  {author} {\bibinfo {author} {\bibfnamefont {F.~P.}\ \bibnamefont
  {Laussy}}, \bibinfo {author} {\bibfnamefont {I.~A.}\ \bibnamefont {Shelykh}},
  \bibinfo {author} {\bibfnamefont {G.}~\bibnamefont {Malpuech}},\ and\
  \bibinfo {author} {\bibfnamefont {A.}~\bibnamefont {Kavokin}},\ }\bibfield
  {title} {\bibinfo {title} {Effects of bose-einstein condensation of exciton
  polaritons in microcavities on the polarization of emitted light},\ }\href
  {https://doi.org/10.1103/PhysRevB.73.035315} {\bibfield  {journal} {\bibinfo
  {journal} {Phys. Rev. B}\ }\textbf {\bibinfo {volume} {73}},\ \bibinfo
  {pages} {035315} (\bibinfo {year} {2006})}\BibitemShut {NoStop}%
\bibitem [{\citenamefont {Rubo}\ \emph {et~al.}(2006)\citenamefont {Rubo},
  \citenamefont {Kavokin},\ and\ \citenamefont {Shelykh}}]{PLARubo}%
  \BibitemOpen
  \bibfield  {author} {\bibinfo {author} {\bibfnamefont {Y.~G.}\ \bibnamefont
  {Rubo}}, \bibinfo {author} {\bibfnamefont {A.~V.}\ \bibnamefont {Kavokin}},\
  and\ \bibinfo {author} {\bibfnamefont {I.~A.}\ \bibnamefont {Shelykh}},\
  }\bibfield  {title} {\bibinfo {title} {Suppression of superfluidity of
  exciton-polaritons by magnetic field},\ }\href
  {https://doi.org/10.1016/j.physleta.2006.05.015} {\bibfield  {journal}
  {\bibinfo  {journal} {Phys. Lett. A}\ }\textbf {\bibinfo {volume} {358}},\
  \bibinfo {pages} {227} (\bibinfo {year} {2006})}\BibitemShut {NoStop}%
\bibitem [{\citenamefont {Larionov}\ \emph {et~al.}(2010)\citenamefont
  {Larionov}, \citenamefont {Kulakovskii}, \citenamefont {H\"ofling},
  \citenamefont {Schneider}, \citenamefont {Worschech},\ and\ \citenamefont
  {Forchel}}]{Larionov2010}%
  \BibitemOpen
  \bibfield  {author} {\bibinfo {author} {\bibfnamefont {A.~V.}\ \bibnamefont
  {Larionov}}, \bibinfo {author} {\bibfnamefont {V.~D.}\ \bibnamefont
  {Kulakovskii}}, \bibinfo {author} {\bibfnamefont {S.}~\bibnamefont
  {H\"ofling}}, \bibinfo {author} {\bibfnamefont {C.}~\bibnamefont
  {Schneider}}, \bibinfo {author} {\bibfnamefont {L.}~\bibnamefont
  {Worschech}},\ and\ \bibinfo {author} {\bibfnamefont {A.}~\bibnamefont
  {Forchel}},\ }\bibfield  {title} {\bibinfo {title} {Polarized nonequilibrium
  {B}ose-{E}instein condensates of spinor exciton polaritons in a magnetic
  field},\ }\href {https://doi.org/10.1103/PhysRevLett.105.256401} {\bibfield
  {journal} {\bibinfo  {journal} {Phys. Rev. Lett.}\ }\textbf {\bibinfo
  {volume} {105}},\ \bibinfo {pages} {256401} (\bibinfo {year}
  {2010})}\BibitemShut {NoStop}%
\bibitem [{\citenamefont {Walker}\ \emph {et~al.}(2011)\citenamefont {Walker},
  \citenamefont {Liew}, \citenamefont {Sarkar}, \citenamefont {Durska},
  \citenamefont {Love}, \citenamefont {Skolnick}, \citenamefont {Roberts},
  \citenamefont {Shelykh}, \citenamefont {Kavokin},\ and\ \citenamefont
  {Krizhanovskii}}]{Walker2011}%
  \BibitemOpen
  \bibfield  {author} {\bibinfo {author} {\bibfnamefont {P.}~\bibnamefont
  {Walker}}, \bibinfo {author} {\bibfnamefont {T.~C.~H.}\ \bibnamefont {Liew}},
  \bibinfo {author} {\bibfnamefont {D.}~\bibnamefont {Sarkar}}, \bibinfo
  {author} {\bibfnamefont {M.}~\bibnamefont {Durska}}, \bibinfo {author}
  {\bibfnamefont {A.~P.~D.}\ \bibnamefont {Love}}, \bibinfo {author}
  {\bibfnamefont {M.~S.}\ \bibnamefont {Skolnick}}, \bibinfo {author}
  {\bibfnamefont {J.~S.}\ \bibnamefont {Roberts}}, \bibinfo {author}
  {\bibfnamefont {I.~A.}\ \bibnamefont {Shelykh}}, \bibinfo {author}
  {\bibfnamefont {A.~V.}\ \bibnamefont {Kavokin}},\ and\ \bibinfo {author}
  {\bibfnamefont {D.~N.}\ \bibnamefont {Krizhanovskii}},\ }\bibfield  {title}
  {\bibinfo {title} {Suppression of {Z}eeman splitting of the energy levels of
  exciton-polariton condensates in semiconductor microcavities in an external
  magnetic field},\ }\href {https://doi.org/10.1103/PhysRevLett.106.257401}
  {\bibfield  {journal} {\bibinfo  {journal} {Phys. Rev. Lett.}\ }\textbf
  {\bibinfo {volume} {106}},\ \bibinfo {pages} {257401} (\bibinfo {year}
  {2011})}\BibitemShut {NoStop}%
\bibitem [{\citenamefont {Fischer}\ \emph {et~al.}(2014)\citenamefont
  {Fischer}, \citenamefont {Brodbeck}, \citenamefont {Chernenko}, \citenamefont
  {Lederer}, \citenamefont {Rahimi-Iman}, \citenamefont {Amthor}, \citenamefont
  {Kulakovskii}, \citenamefont {Worschech}, \citenamefont {Kamp}, \citenamefont
  {Durnev}, \citenamefont {Schneider}, \citenamefont {Kavokin},\ and\
  \citenamefont {H\"ofling}}]{Fischer2014}%
  \BibitemOpen
  \bibfield  {author} {\bibinfo {author} {\bibfnamefont {J.}~\bibnamefont
  {Fischer}}, \bibinfo {author} {\bibfnamefont {S.}~\bibnamefont {Brodbeck}},
  \bibinfo {author} {\bibfnamefont {A.~V.}\ \bibnamefont {Chernenko}}, \bibinfo
  {author} {\bibfnamefont {I.}~\bibnamefont {Lederer}}, \bibinfo {author}
  {\bibfnamefont {A.}~\bibnamefont {Rahimi-Iman}}, \bibinfo {author}
  {\bibfnamefont {M.}~\bibnamefont {Amthor}}, \bibinfo {author} {\bibfnamefont
  {V.~D.}\ \bibnamefont {Kulakovskii}}, \bibinfo {author} {\bibfnamefont
  {L.}~\bibnamefont {Worschech}}, \bibinfo {author} {\bibfnamefont
  {M.}~\bibnamefont {Kamp}}, \bibinfo {author} {\bibfnamefont {M.}~\bibnamefont
  {Durnev}}, \bibinfo {author} {\bibfnamefont {C.}~\bibnamefont {Schneider}},
  \bibinfo {author} {\bibfnamefont {A.~V.}\ \bibnamefont {Kavokin}},\ and\
  \bibinfo {author} {\bibfnamefont {S.}~\bibnamefont {H\"ofling}},\ }\bibfield
  {title} {\bibinfo {title} {Anomalies of a nonequilibrium spinor polariton
  condensate in a magnetic field},\ }\href
  {https://doi.org/10.1103/PhysRevLett.112.093902} {\bibfield  {journal}
  {\bibinfo  {journal} {Phys. Rev. Lett.}\ }\textbf {\bibinfo {volume} {112}},\
  \bibinfo {pages} {093902} (\bibinfo {year} {2014})}\BibitemShut {NoStop}%
\bibitem [{\citenamefont {Kudlis}\ \emph {et~al.}(2024)\citenamefont {Kudlis},
  \citenamefont {Novokreschenov},\ and\ \citenamefont {Shelykh}}]{kudlis2024}%
  \BibitemOpen
  \bibfield  {author} {\bibinfo {author} {\bibfnamefont {A.}~\bibnamefont
  {Kudlis}}, \bibinfo {author} {\bibfnamefont {D.}~\bibnamefont
  {Novokreschenov}},\ and\ \bibinfo {author} {\bibfnamefont {I.~A.}\
  \bibnamefont {Shelykh}},\ }\href {https://arxiv.org/abs/2412.09245} {\bibinfo
  {title} {Extended {XY} model for spinor polariton simulators}} (\bibinfo
  {year} {2024}),\ \Eprint {https://arxiv.org/abs/2412.09245} {arXiv:2412.09245
  [cond-mat.mes-hall]} \BibitemShut {NoStop}%
\bibitem [{\citenamefont {Berloff}\ \emph {et~al.}(2017)\citenamefont
  {Berloff}, \citenamefont {Silva}, \citenamefont {Kalinin}, \citenamefont
  {Askitopoulos}, \citenamefont {Töpfer}, \citenamefont {Cilibrizzi},
  \citenamefont {Langbein},\ and\ \citenamefont {Lagoudakis}}]{Berloff2017}%
  \BibitemOpen
  \bibfield  {author} {\bibinfo {author} {\bibfnamefont {N.~G.}\ \bibnamefont
  {Berloff}}, \bibinfo {author} {\bibfnamefont {M.}~\bibnamefont {Silva}},
  \bibinfo {author} {\bibfnamefont {K.}~\bibnamefont {Kalinin}}, \bibinfo
  {author} {\bibfnamefont {A.}~\bibnamefont {Askitopoulos}}, \bibinfo {author}
  {\bibfnamefont {J.~D.}\ \bibnamefont {Töpfer}}, \bibinfo {author}
  {\bibfnamefont {P.}~\bibnamefont {Cilibrizzi}}, \bibinfo {author}
  {\bibfnamefont {W.}~\bibnamefont {Langbein}},\ and\ \bibinfo {author}
  {\bibfnamefont {P.~G.}\ \bibnamefont {Lagoudakis}},\ }\bibfield  {title}
  {\bibinfo {title} {Realizing the classical {XY} {H}amiltonian in polariton
  simulators},\ }\href {https://doi.org/10.1038/nmat4971} {\bibfield  {journal}
  {\bibinfo  {journal} {Nat. Mater.}\ }\textbf {\bibinfo {volume} {16}},\
  \bibinfo {pages} {1120} (\bibinfo {year} {2017})}\BibitemShut {NoStop}%
\bibitem [{\citenamefont {Shelykh}\ \emph {et~al.}(2007)\citenamefont
  {Shelykh}, \citenamefont {Rubo},\ and\ \citenamefont
  {Kavokin}}]{shelykh2007SaM}%
  \BibitemOpen
  \bibfield  {author} {\bibinfo {author} {\bibfnamefont {I.}~\bibnamefont
  {Shelykh}}, \bibinfo {author} {\bibfnamefont {Y.}~\bibnamefont {Rubo}},\ and\
  \bibinfo {author} {\bibfnamefont {A.}~\bibnamefont {Kavokin}},\ }\bibfield
  {title} {\bibinfo {title} {Renormalized dispersion of elementary excitations
  in spinor polariton condensates},\ }\href
  {https://doi.org/https://doi.org/10.1016/j.spmi.2007.03.006} {\bibfield
  {journal} {\bibinfo  {journal} {Superlattices Microstruct.}\ }\textbf
  {\bibinfo {volume} {41}},\ \bibinfo {pages} {313} (\bibinfo {year} {2007})},\
  \bibinfo {note} {proceedings of the 6th International Conference on Physics
  of Light-Matter Coupling in Nanostructures}\BibitemShut {NoStop}%
\bibitem [{\citenamefont {Brtka}\ \emph {et~al.}(2010)\citenamefont {Brtka},
  \citenamefont {Gammal},\ and\ \citenamefont {Malomed}}]{Brtka2010}%
  \BibitemOpen
  \bibfield  {author} {\bibinfo {author} {\bibfnamefont {M.}~\bibnamefont
  {Brtka}}, \bibinfo {author} {\bibfnamefont {A.}~\bibnamefont {Gammal}},\ and\
  \bibinfo {author} {\bibfnamefont {B.~A.}\ \bibnamefont {Malomed}},\
  }\bibfield  {title} {\bibinfo {title} {Hidden vorticity in binary
  {B}ose-{E}instein condensates},\ }\href
  {https://doi.org/10.1103/PhysRevA.82.053610} {\bibfield  {journal} {\bibinfo
  {journal} {Phys. Rev. A}\ }\textbf {\bibinfo {volume} {82}},\ \bibinfo
  {pages} {053610} (\bibinfo {year} {2010})}\BibitemShut {NoStop}%
\bibitem [{\citenamefont {Yulin}\ \emph {et~al.}(2020)\citenamefont {Yulin},
  \citenamefont {Nalitov},\ and\ \citenamefont {Shelykh}}]{Yulin2020}%
  \BibitemOpen
  \bibfield  {author} {\bibinfo {author} {\bibfnamefont {A.~V.}\ \bibnamefont
  {Yulin}}, \bibinfo {author} {\bibfnamefont {A.~V.}\ \bibnamefont {Nalitov}},\
  and\ \bibinfo {author} {\bibfnamefont {I.~A.}\ \bibnamefont {Shelykh}},\
  }\bibfield  {title} {\bibinfo {title} {Spinning polariton vortices with
  magnetic field},\ }\href {https://doi.org/10.1103/PhysRevB.101.104308}
  {\bibfield  {journal} {\bibinfo  {journal} {Phys. Rev. B}\ }\textbf {\bibinfo
  {volume} {101}},\ \bibinfo {pages} {104308} (\bibinfo {year}
  {2020})}\BibitemShut {NoStop}%
\bibitem [{\citenamefont {Rubo}(2022)}]{Rubo2022}%
  \BibitemOpen
  \bibfield  {author} {\bibinfo {author} {\bibfnamefont {Y.~G.}\ \bibnamefont
  {Rubo}},\ }\bibfield  {title} {\bibinfo {title} {Spin-orbital effect on the
  polariton state in traps},\ }\href
  {https://doi.org/10.1103/PhysRevB.106.235306} {\bibfield  {journal} {\bibinfo
   {journal} {Phys. Rev. B}\ }\textbf {\bibinfo {volume} {106}},\ \bibinfo
  {pages} {235306} (\bibinfo {year} {2022})}\BibitemShut {NoStop}%
\bibitem [{\citenamefont {Chestnov}\ \emph {et~al.}(2023)\citenamefont
  {Chestnov}, \citenamefont {Kondratenko}, \citenamefont {Demirchyan},\ and\
  \citenamefont {Kavokin}}]{Chestnov2023}%
  \BibitemOpen
  \bibfield  {author} {\bibinfo {author} {\bibfnamefont {I.}~\bibnamefont
  {Chestnov}}, \bibinfo {author} {\bibfnamefont {K.}~\bibnamefont
  {Kondratenko}}, \bibinfo {author} {\bibfnamefont {S.}~\bibnamefont
  {Demirchyan}},\ and\ \bibinfo {author} {\bibfnamefont {A.}~\bibnamefont
  {Kavokin}},\ }\bibfield  {title} {\bibinfo {title} {Symmetry breaking and
  superfluid currents in a split-ring spinor polariton condensate},\ }\href
  {https://doi.org/10.1103/PhysRevB.107.245302} {\bibfield  {journal} {\bibinfo
   {journal} {Phys. Rev. B}\ }\textbf {\bibinfo {volume} {107}},\ \bibinfo
  {pages} {245302} (\bibinfo {year} {2023})}\BibitemShut {NoStop}%
\bibitem [{\citenamefont {Rubo}(2007)}]{Rubo2007PRL}%
  \BibitemOpen
  \bibfield  {author} {\bibinfo {author} {\bibfnamefont {Y.~G.}\ \bibnamefont
  {Rubo}},\ }\bibfield  {title} {\bibinfo {title} {Half vortices in exciton
  polariton condensates},\ }\href
  {https://doi.org/10.1103/PhysRevLett.99.106401} {\bibfield  {journal}
  {\bibinfo  {journal} {Phys. Rev. Lett.}\ }\textbf {\bibinfo {volume} {99}},\
  \bibinfo {pages} {106401} (\bibinfo {year} {2007})}\BibitemShut {NoStop}%
\bibitem [{\citenamefont {Flayac}\ \emph {et~al.}(2010)\citenamefont {Flayac},
  \citenamefont {Shelykh}, \citenamefont {Solnyshkov},\ and\ \citenamefont
  {Malpuech}}]{Flayac2010}%
  \BibitemOpen
  \bibfield  {author} {\bibinfo {author} {\bibfnamefont {H.}~\bibnamefont
  {Flayac}}, \bibinfo {author} {\bibfnamefont {I.~A.}\ \bibnamefont {Shelykh}},
  \bibinfo {author} {\bibfnamefont {D.~D.}\ \bibnamefont {Solnyshkov}},\ and\
  \bibinfo {author} {\bibfnamefont {G.}~\bibnamefont {Malpuech}},\ }\bibfield
  {title} {\bibinfo {title} {Topological stability of the half-vortices in
  spinor exciton-polariton condensates},\ }\href
  {https://doi.org/10.1103/PhysRevB.81.045318} {\bibfield  {journal} {\bibinfo
  {journal} {Phys. Rev. B}\ }\textbf {\bibinfo {volume} {81}},\ \bibinfo
  {pages} {045318} (\bibinfo {year} {2010})}\BibitemShut {NoStop}%
\bibitem [{\citenamefont {Król}\ \emph {et~al.}(2019)\citenamefont {Król},
  \citenamefont {Mirek}, \citenamefont {Stephan}, \citenamefont {Lekenta},
  \citenamefont {Rousset}, \citenamefont {Pacuski}, \citenamefont {Kavokin},
  \citenamefont {Matuszewski}, \citenamefont {Szczytko},\ and\ \citenamefont
  {Piętka}}]{Krol2019}%
  \BibitemOpen
  \bibfield  {author} {\bibinfo {author} {\bibfnamefont {M.}~\bibnamefont
  {Król}}, \bibinfo {author} {\bibfnamefont {R.}~\bibnamefont {Mirek}},
  \bibinfo {author} {\bibfnamefont {D.}~\bibnamefont {Stephan}}, \bibinfo
  {author} {\bibfnamefont {K.}~\bibnamefont {Lekenta}}, \bibinfo {author}
  {\bibfnamefont {J.-G.}\ \bibnamefont {Rousset}}, \bibinfo {author}
  {\bibfnamefont {W.}~\bibnamefont {Pacuski}}, \bibinfo {author} {\bibfnamefont
  {A.~V.}\ \bibnamefont {Kavokin}}, \bibinfo {author} {\bibfnamefont
  {M.}~\bibnamefont {Matuszewski}}, \bibinfo {author} {\bibfnamefont
  {J.}~\bibnamefont {Szczytko}},\ and\ \bibinfo {author} {\bibfnamefont
  {B.}~\bibnamefont {Piętka}},\ }\bibfield  {title} {\bibinfo {title} {Giant
  spin meissner effect in a nonequilibrium exciton-polariton gas},\ }\href
  {https://doi.org/10.1103/PhysRevB.99.115318} {\bibfield  {journal} {\bibinfo
  {journal} {Phys. Rev. B}\ }\textbf {\bibinfo {volume} {99}},\ \bibinfo
  {pages} {115318} (\bibinfo {year} {2019})}\BibitemShut {NoStop}%
\bibitem [{\citenamefont {Sawicki}\ \emph {et~al.}(2024)\citenamefont
  {Sawicki}, \citenamefont {Dovzhenko}, \citenamefont {Wang}, \citenamefont
  {Cookson}, \citenamefont {Sigurðsson},\ and\ \citenamefont
  {Lagoudakis}}]{Sawicki2024}%
  \BibitemOpen
  \bibfield  {author} {\bibinfo {author} {\bibfnamefont {K.}~\bibnamefont
  {Sawicki}}, \bibinfo {author} {\bibfnamefont {D.}~\bibnamefont {Dovzhenko}},
  \bibinfo {author} {\bibfnamefont {Y.}~\bibnamefont {Wang}}, \bibinfo {author}
  {\bibfnamefont {T.}~\bibnamefont {Cookson}}, \bibinfo {author} {\bibfnamefont
  {H.}~\bibnamefont {Sigurðsson}},\ and\ \bibinfo {author} {\bibfnamefont
  {P.~G.}\ \bibnamefont {Lagoudakis}},\ }\bibfield  {title} {\bibinfo {title}
  {Occupancy-driven zeeman suppression and inversion in trapped polariton
  condensates},\ }\href {https://doi.org/10.1103/PhysRevB.109.125307}
  {\bibfield  {journal} {\bibinfo  {journal} {Phys. Rev. B}\ }\textbf {\bibinfo
  {volume} {109}},\ \bibinfo {pages} {125307} (\bibinfo {year}
  {2024})}\BibitemShut {NoStop}%
\bibitem [{\citenamefont {Bochin}\ \emph {et~al.}(2024)\citenamefont {Bochin},
  \citenamefont {Chestnov},\ and\ \citenamefont {Nalitov}}]{Bochin2024}%
  \BibitemOpen
  \bibfield  {author} {\bibinfo {author} {\bibfnamefont {A.}~\bibnamefont
  {Bochin}}, \bibinfo {author} {\bibfnamefont {I.}~\bibnamefont {Chestnov}},\
  and\ \bibinfo {author} {\bibfnamefont {A.}~\bibnamefont {Nalitov}},\
  }\bibfield  {title} {\bibinfo {title} {Giant effective g-factor due to spin
  bifurcations in polariton condensates},\ }\href
  {https://doi.org/10.1134/S0021364024601349} {\bibfield  {journal} {\bibinfo
  {journal} {JETP Letters}\ }\textbf {\bibinfo {volume} {119}},\ \bibinfo
  {pages} {917} (\bibinfo {year} {2024})}\BibitemShut {NoStop}%
\end{thebibliography}
%

\end{document}